\documentclass[preprint]{elsarticle}
\usepackage[english]{babel}
\usepackage[utf8x]{inputenc}
\usepackage[T1]{fontenc}
\usepackage{floatrow}
\usepackage{trackchanges}
\journal{Network Neuroscience}

\usepackage[a4paper,top=2.5cm,bottom=2.5cm,left=2.5cm,right=2.5cm,marginparwidth=1.75cm]{geometry}

\usepackage{amsmath}
\usepackage{graphicx}
\usepackage[colorinlistoftodos]{todonotes}
\usepackage[colorlinks=true, allcolors=blue]{hyperref}
\usepackage{framed}

\begin{document}

\begin{frontmatter}

\title{Disentangling causal webs in the~brain using functional Magnetic Resonance Imaging: A review of current approaches}
\date{\vspace{-5ex}}

%% Group authors per affiliation:
\author[DCCNAdress,RadboudAddress]{Natalia Z. Bielczyk\corref{mycorrespondingauthor}}
\cortext[mycorrespondingauthor]{Corresponding author}
\ead{natalia.bielczyk@gmail.com}
\author[DCCNAdress,ChariteAddress]{Sebo Uithol}
\author[DCCNAdress,RadboudAddress]{Tim van Mourik}
\author[DCCNAdress,RadboudAddress2]{Paul Anderson}
\author[DCCNAdress,RadboudAddress]{Jeffrey C. Glennon}
\author[DCCNAdress,RadboudAddress]{Jan K. Buitelaar}

\address[DCCNAdress]{Donders Institute for Brain, Cognition and Behavior, Kapittelweg 29, 6525 EN Nijmegen, the Netherlands}
\address[RadboudAddress]{Radboud University Nijmegen Medical Centre, Geert Grooteplein Zuid 10, 6525 GA Nijmegen, the Netherlands}
\address[RadboudAddress2]{Radboud University Nijmegen, Comeniuslaan 4, 6525 HP Nijmegen, the Netherlands}
\address[ChariteAddress]{Charit\'{e} Universit\"{a}tsmedizin, Bernstein Centre for Computational Neuroscience, Philippstrasse 13, haus 6, 10119, Berlin, Germany}

\begin{abstract}
In the past two decades, functional Magnetic Resonance Imaging has been used to relate neuronal network activity to cognitive processing and behaviour. Recently this approach has been augmented by algorithms that allow us to infer causal links between component populations of neuronal networks. Multiple inference procedures have been proposed to approach this research question but so far, each method has limitations when it comes to establishing whole-brain connectivity patterns. 
In this paper, we discuss eight ways to infer causality in fMRI research: Bayesian Nets, Dynamical Causal Modelling, Granger Causality, Likelihood Ratios, LiNGAM, Patel's Tau, Structural Equation Modelling, and Transfer Entropy. We finish with formulating some recommendations for the~future directions in this area.
\end{abstract}


\end{frontmatter}

\section{Introduction}\label{sec:intro}
\subsection{What is causality?}
Although inferring causal relations is a~fundamental aspect of scientific research, the notion of causation itself is notoriously difficult to define. The basic idea is straightforward: When process A~is the cause of process B, A~is necessarily in the past from B, and without A, B would not occur. But in practice, and in dynamic systems such as the brain in particular, the picture is far less clear. First, for any event a~large number of (potential) causes can be identified. The efficacy of certain neuronal process in producing behavior is dependent on the state of many other (neuronal) processes, but also on the~availability of glucose and oxygen in the brain, etc. In a~neuroscientific context, we are generally not interested in most of these causes, but only in a~cause that stands out in such a~way that it is deemed to provide a~substantial part of the explanation, for instance causes that vary with the experimental conditions. However, the contrast between relevant and irrelevant causes (in terms of explanatory power) is arbitrary and strongly dependent on experimental setup, contextual factors, etc. For instance, respiratory movement is typically considered a~confound in fMRI experiments, unless the research question concerns the~influence of respiration speed on the~dynamics of the~neuronal networks. 

In dynamic systems, causal processes are unlikely to be part of a~unidirectional chain of events, but rather a~causal web, with often mutual influences between process A~and B~\cite{mannino2015}. As a~result, it is hard to maintain the temporal ordering of cause and effect and, indeed, a~clear separation between them~\cite{schurger2015}.

Furthermore, causation can never be observed directly, just correlation~\cite{hume1772}. When a~correlation is highly stable, we are inclined to infer a~causal link. Additional information is then needed to assess the direction of the assumed causal link, as correlation indicates for association and not for causation~\cite{altman2015}. For example, the motor cortex is always active when a~movement is made, so we assume a~causal link between the two phenomena. The anatomical and physiological properties of the motor cortex, and the timing of the two phenomena provide clues about the direction of causality (i.e. cortical activity causes the movement, and not the other way around). However, only intervention studies, such as delivering Transcranial Magnetic Stimulation (TMS,~\cite{kim2009}) pulses over the motor cortex or lesion studies, can confirm the~causal link between the~activity in the~motor cortex and behavior. %(but technically, also here only a~correlation between the pulse and the muscle twitch can be established). 

Causal studies in fMRI are based on three types of correlations: correlating neuronal activity to 1) mental and behavioral phenomena; 2) to physiological states (such as neurotransmitters, hormones, etc.), and 3) to neuronal activity in other parts of the brain. In this review we will focus on the last field of research: establishing causal connections between activity in two or more brain areas.
\subsection{A~note on the limitations of fMRI data}
\label{sec:limitations}
fMRI studies currently use a variety of algorithms to infer causal links~\cite{fornito2013,smith2011}. All these methods have different assumptions, advantages and disadvantages (see for instance \cite{valdes-sosa2011,stephan2012}), and approach the~problem from different angles. An important reason for this variety of approaches is the~complex nature of fMRI data, which imposes severe restrictions on the possibility of finding causal relations using MRI.

\subsubsection{Temporal resolution and haemodynamics}\label{sec:limitations_hemod}
First, and best known, the~temporal resolution of the~image acquisition in MR imaging is generally restricted to a sampling rate $<1[Hz]$. Recently, multiband fMRI protocols have gained in popularity~\cite{feinberg2013}, which increases the upper limit for the scanning frequency to up to $10[Hz]$, albeit at the cost of a severely decreased signal-to-noise ratio. However, no imaging protocol (including multiband imaging) can overcome the limitation of the recorded signal itself: the~lagged change in blood oxygenation, which peaks 3 to 6 seconds after neuronal firing in the~adult human brain~\cite{arichi2012}. The haemodynamic response thus acts as a~low-pass filter, which results in high correlations between activity in consecutive frames~\cite{ramsey2010}. Since the haemodynamic lags (understood as the peaks of the haemodynamic response) are region- and subject- specific~\cite{devonshire2012} and vary over time~\cite{glomb2017a}, it is difficult to infer causality between two time series with potentially different haemodynamic lags~\cite{bielczyk2017}. Computational work by Seth et al.~\cite{seth2013} suggests that upsampling the~signal to low TRs ($<0.1[s]$) might potentially overcome this issue. Furthermore, haemodynamics typically fluctuates in time. These slow fluctuations, similarly to other low frequency artifacts such as heartbeat or body movements, should be removed from the~datasets through high-pass filtering before the~inference procedure~\cite{ramsey2014}. 

\subsubsection{Signal-to-noise ratio}\label{sec:limitations_snr}
Second, fMRI data is characterized by a relatively low signal-to-noise ratio. In grey matter, the recorded haemodynamic response changes by 1-2\% at field strengths of $1.5-2.0[T]$ (\cite{ogawa1993, boxerman1995}), and by 5-6\% at field strengths of $4.0[T]$. Moreover, typical fMRI protocols generate relatively short time series. For example, the new Human Connectome Project resting state datasets~\cite{vanessen2013} do not contain more than a~few hundred to maximally few thousand samples. Two most popular ways of improving on the~SNR in fMRI datasets are averaging signals over multiple voxels~\cite{friston_book} and spatial smoothing~\cite{triantafyllou2006}.%The~short BOLD time series limits the option of improving signal-to-noise ratios through averaging across samples. `

\subsubsection{Caveats associated with region definition}\label{sec:limitations_regiondef}
Third, in order to propose a~causal model, one first needs to define the~nodes of the~network. A~single voxel does not represent a~biologically meaningful part of the~brain~\cite{stanley2013}. Therefore, before attempting to establish causal connection in the network, one needs to integrate the~BOLD time series over regions of interest (ROIs): groups of voxels that are assumed to share a common signal with a neuroscientific meaning. Choosing the~optimal regions of interest for a study is a~complex problem~\cite{poldrack2007,marrelec2011,thirion2014,fornito2013,kelly2012}. In task-based fMRI, ROIs are often chosen on the basis of activation patterns revealed by the standard GLM analysis~\cite{friston_book}. 

On the other hand, in research into resting state brain activity, the analysis is usually exploratory and the~connectivity in larger, meso- and macroscale networks is typically considered. In that case, a few strategies to ROI definition are possible. First, one can define regions of interest on the~basis of brain anatomy. However, a consequence of this strategy could be that BOLD activity related to the cognitive process of interest will be mixed with other, unrelated activity within the ROIs. This is particularly likely to happen given that brain structure is not exactly replicable across individuals, so that a specific area cannot be defined reliably based on location alone. As indicated in the~computational study by Smith et al.~\cite{smith2011}, and also in a~recent study by Bielczyk et al.~\cite{bielczyk2017}, such signal mixing is detrimental to causal inference and causes all the existing methods for causal inference in fMRI to underperform. What these studies demonstrate is that parcellating into ROIs based on anatomy rather than common activity, can induce additional scale-free background noise in the~networks. Since this noise has high power in low frequencies, the~modelled BOLD response cannot effectively filter it out. As a~consequence, the~signatures of different connectivity patterns are getting lost.

As an~alternative to anatomical parcellation, choosing ROIs can be performed in a~functional, data-driven fashion. There are multiple techniques developed to reach this goal, and to list some recent examples: Instantaneous Correlations Parcellation implemented through a~hierarchical Independent Component Analysis (ICP,~\cite{vanoort2017}), probabilistic parcellation based on Chinese restaurant process~\cite{janssen2015}, graph clustering based on inter-voxel correlations~\cite{vanheuvel2008}, large-scale network identification through comparison between correlations among ROIs versus a~model of the correlations generated by the~noise (LSNI,~\cite{bellec2006}), multi-level bootstrap analysis~\cite{bellec2010}, clustering of voxels revealing common causal patterns in terms of Granger Causality~\cite{dsouza2017}, spatially constrained hierarchical clustering~\cite{blumensath2013} and algorithms providing a~trade-off between machine learning techniques and knowledge coming from neuroanatomy~\cite{glasser2016}. 

Another possibility to reduce the effect of mixing signals is to perform Principal Component Analysis (PCA,~\cite{jolliffe2002,schlens2014b}) and separate the~BOLD time series within each anatomical region into a~sum of orthogonal signals (eigenvariates) and choose only the~signal with the~highest contribution to the BOLD signal (the~first eigenvariate,~\cite{friston2003}), instead of averaging activity over full anatomical regions. Finally, one can build ROIs on the basis of patterns of activation only (task localizers~\cite{fedorenko2010, heinzle2012}). However, this approach cannot be applied to resting state research. In this work, we assume that the definition of ROIs has been performed by the~researcher prior to the~causal inference, and we do not discuss it any further. 

\subsection{Criteria for evaluating methods for causal inference in functional Magnetic Resonance Imaging}~\label{sec:criteria}
Given the~aforementioned characteristics of fMRI data (low temporal resolution, slow haemodynamics, low signal-to-noise ratio) and the fact that causal webs in the brain are likely dense and dynamic, is it in principle possible to investigate causality in the brain using MRI? Multiple distinct families of models have been developed in order to approach this problem over the~past two decades. One can look at the~methods from different angles, and classify them into different categories. 

One important distinction proposed by Friston et al.~\cite{friston2013}, includes division of methods with respect to \textit{the~depth of the~neuroimaging measurements at which a~method is defined}. Most methods (such as the~original formulation of Structural Equation Modeling for fMRI~\cite{mcintosh1994}, see: chapter~\ref{sec:sem}) operate on the~experimental \textit{observables}, i.e. the~measured BOLD responses. These methods are referred to as~\textit{directed functional connectivity} measures. On the~contrary, other methods (e.g., Dynamic Causal Modeling, see: chapter~\ref{sec:dcm}) consider the~underlying neuronal processes. These methods are referred to as~\textit{effective connectivity} measures. Mind that while some methods such as Dynamic Causal Modeling are hardwired to assess effective connectivity (as they are built upon a~generative model), other methods can be used both as a~method to assess directed functional connectivity or effective connectivity. E.g., in Granger Causality research (see: chapter~\ref{sec:gc}), a~blind deconvolution in often used in order to deconvolve the~observed BOLD responses into an~underlying neuronal time series~\cite{david2008,ryali2011,ryali2016,hutcheson2015,wheelock2014,sathian2013,goodyear2016}, which allows for assessing effective connectivity. On the~contrary, when Granger Causality is used without deconvolution~\cite{zhao2016,regner2016,chen2017}, it is a~directed functional connectivity method. Of course, both scenarios have pros and cons, as blind deconvolution can be a~very noisy operation~\cite{bush2015} and for more details, please see Friston et al.~\cite{friston2013}.

Another important distinction was proposed by Valdes-Sosa et al.~\cite{valdes-sosa2011}. According to this point of view, methods can be divided on the~basis of the~\textit{approach towards temporal sequence of the~samples}: some of the~methods are based on the~temporal sequence of the~signals (e.g. TE or GC), or rely on the~dynamics expressed by state-space equations (so-called \textit{state-space} models, e.g., DCM), while other methods do not draw information from the~sequence in time, and solely focus on the~statistical properties of the~time series (so-called \textit{structural models}, e.g. BNs).

In this work, we would like to propose another classification of methods for causal inference in fMRI. First, we identify nine characteristics of models used to study causality. Then, we compare and contrast the popular approaches to the~causal research in fMRI according to these criteria. Our list of features of causality is as follows:
\begin{enumerate}
\item \textit{Sign of connections}: Can the method distinguish between excitatory and inhibitory causal relations? In this context, we do not mean \textit{synaptic} effects, but rather an overall driving or attenuating impact of the~activity in one brain region on the~activity in another region. Certain methods only detect the~existence of causal influence from the~BOLD responses, whereas others can distinguish between these distinct forms of influence.
\item \textit{Strength of connections}: Can the method distinguish between weak and strong connections, apart from indicating the directionality of connections at a~certain confidence level?
\item \textit{Confidence intervals}: How are the~confidence intervals for the~connections determined?
\item \textit{Bidirectionality}: Can the method pick up bidirectional connections $X \rightleftharpoons Y$, or only indicate the strongest of the two connections $X \rightarrow Y$ and $Y \rightarrow X$? Some methods do not \textit{allow} for bidirectional relations, since they cannot deal with cycles in the network.
\item \textit{Immediacy}: Does the method specifically identify direct influences $X \rightarrow Y$, or does it pool across direct and indirect influences $Z_i$: $X \rightarrow Z_i \rightarrow Y$? We assume that $Z_i$ represent nodes in the~network, and the~activity in these nodes is measured (otherwise $Z_i$ become a~latent confounder). While some methods aim to make this distinction, others highlight any influence $X\rightarrow Y$, whenever it is direct or not. 
\item \textit{Resilience to confounds}: Does the method correct for possible spurious causal effects from a~common source ($Z \rightarrow X$, $Z \rightarrow Y$, so we infer $X \rightarrow Y$ and/or $Y \rightarrow X$), or other confounders? In general, confounding variables are an~issue to all the~methods for causal inference, especially when a~given study is non-interventional~\cite{rohrer2017}, however different methods can suffer from these issues to a~different extent.
\item \textit{Type of inference}: Does the method probe causality through classical hypothesis testing or through model comparison? Hypothesis-based methods will test a~null hypothesis $H_0$ that there is no~causal link between two variables, against a~hypothesis $H_1$ that there is causal link between the~two. In contrast, model-comparison-based methods do not have an~explicit null hypothesis. Instead, evidence for a~predefined set of models is computed. In particular cases, when the~investigated network contains only a~few nodes and the~estimation procedure is computationally cheap, a search through all the~connectivity patterns by means of model comparison is possible. In all the~other cases, prior knowledge is necessary to select a~subset of possible models for model comparison.
\item \textit{Computational cost}: What is the computational complexity of the inference procedure? In the case of model comparison, the~computational cost refers to the~cost of finding the likelihood of a~single model, as the~range of possible models depends on the research question. This can lead to practical limitations based on computing power.
\item \textit{Size of the network}: What sizes of network does the method allow for? Some methods are restricted in the number of nodes that it allows, for computational or interpretational reasons.
\end{enumerate}

In certain applications, an~additional criterion of \textit{empirical accuracy in realistic simulation} could be of help to evaluate the~method. Testing the~method on synthetic, ground truth datasets available for the~research problem at hand can give a~good picture on whether or not the~method gives reliable results when applied to experimental datasets. In fMRI research, multiple methods for causal inference were directly compared to each other in a~the~seminal simulation study by Smith et al. In this study, the~authors employed a~Dynamic Causal Modeling generative model~\cite{friston2003}, introduced in Section~\ref{sec:dcm} in order to create synthetic datasets with a~known ground truth. Surprisingly, most of the methods struggled to perform above chance level, even though the test networks were sparse and the noise levels introduced to the model were low compared to what one would expect in real recordings. In this manuscript, we will refer to this study throughout the~text. However, we will not list empirical accuracy as a~separate criterion, for two reasons. Firstly, some of the~methods reviewed here, e.g. SEM, were not tested on the~synthetic benchmark datasets. Secondly, the~most popular method in the~field, Dynamic Causal Modeling~\cite{friston2003}, builds on the~same generative model that is used for comparing methods to each other in Smith's stud. Therefore, it is hard to make a~fair comparison between DCM and other methods in the~field using this generative model.

In the~following chapters, the~references to this~'causality list' will be marked in the~text with lowercase indices.

\begin{figure}[h]
\begin{framed}
\centering
\includegraphics[width=0.85\textwidth]{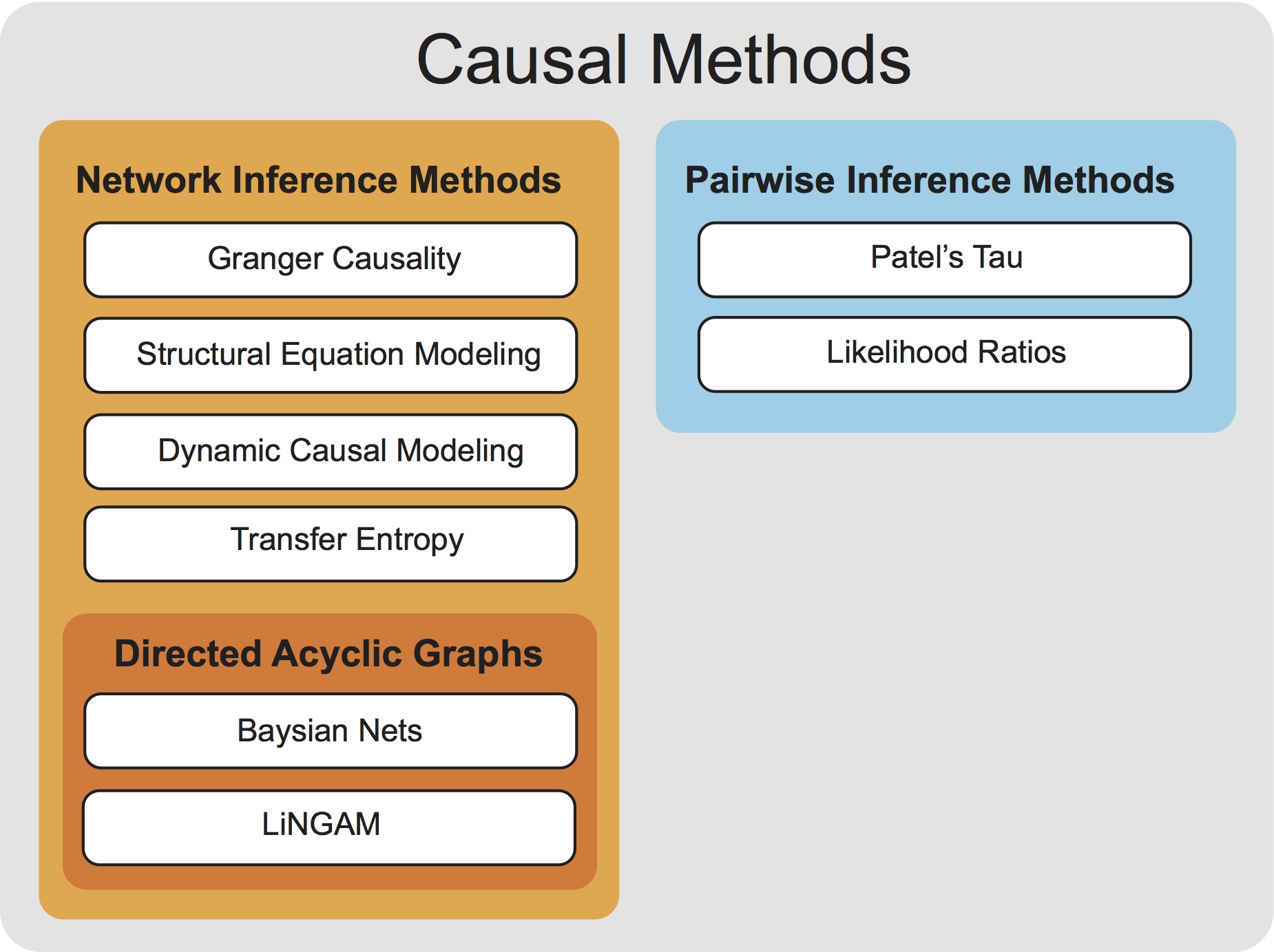}
\end{framed}
\caption{Causal research in functional Magnetic Resonance Imaging. The discussed methods can be divided into two families: Network Inference Methods, which are based on a one-step multivariate procedure, and Pairwise Inference Methods, which are based on a two-step pairwise inference procedures. As pairwise methods by definition establish causal connections on a node-by-node basis, the network as a whole cannot be guaranteed to be of any particular structure.}
\label{fig:fig1}
\end{figure}

With respect to assumptions made on the connectivity structure, the approaches discussed here can be divided into three main groups (Fig.~\ref{fig:fig1}). The~first group comprises multivariate methods that search for directed graphs without imposing any particular structure onto the~graph: Granger Causality~\cite{seth2015}, Transfer Entropy~\cite{marrelec2006}, Structural Equation Modeling~\cite{mcintosh1994} and Dynamic Causal Modeling~\cite{friston2003}. These methods will be referred to as network-wise models throughout the manuscript. The~second group of methods is also multivariate, but requires an~additional assumption of acyclicity. Models in this group assume that information travels through the~brain by~feed-forward projections only. As a result, the network can always be represented by a~Directed Acyclic Graph (DAG,~\cite{thulasiraman1992}). Methods in this group include Linear Non-Gaussian Acyclic Models (LiNGAM,~\cite{shimizu2006}) and Bayesian Nets~\cite{mumford2014}, and will be referred to as hierarchical network-wise models throughout the manuscript. The~last group of methods, referred to as pairwise methods, use a two-stage procedure: first, a~map of nondirectional functional connections is rendered, and second, the directionality in each connection is assessed. Since these methods focus on pairwise connections rather than complete network architectures, they by definition do not impose network assumptions like acyclicity. Patel's tau~\cite{patel2006} and Pairwise Likelihood Ratios~\cite{hyvarinen2013} are members of this group. In this review, we do not include studying a~coupling between brain region and the~rest of the~brain with relation to a~particular cognitive task, The~Psycho-Physiological Interactions (PPIs~\cite{friston1997}), as we are only focused on the~methods for assessing causal links within brain networks, and we do not include brain-behavior causal interactions.

% the separate methods:

% [a] network level + non-DAG:
\section{Network-wise methods}
The~first group of models that we discuss in this review involves multivariate methods: methods that simultaneously assess all causal links in the~network - specifically, Granger Causality~\cite{granger1969}, Transfer Entropy~\cite{schreiber2000}, Structural Equation Modeling~\cite{wright1920} and Dynamic Causal Modeling~\cite{friston2003}. These methods do not pose any constraints on the~connectivity structure. Granger Causality, Transfer Entropy and Structural Equation Modeling infer causal processes through classical hypothesis testing. As there are no limits to the~size of the analyzed network, these methods allow for (relatively) hypothesis-free discovery. Dynamic Causal Modeling on the~other hand, compares  a~number of predefined causal structures in networks of only a~few nodes. As such, it requires a~specific hypothesis based on prior knowledge.
\subsection{Granger causality}\label{sec:gc}
%Introducing the idea
Clive Granger introduced Granger Causality (GC) in the~field of economics~\cite{granger1969}. GC has found its way into many other disciplines, including fMRI research~\cite{roebroeck2011,bressler2011,seth2015,solo2016}. GC is based on prediction~\cite{Diebold2001}: the~signal in a~certain region is dependent on its past values. Therefore, a~time series $Y(t)$ at time point $t$ can be partly predicted by its past values $Y(t-i)$. A~signal in an upstream region is followed by the~same signal in a downstream region with a~certain temporal lag. Therefore, if prediction of $Y(t)$ improves when past values of another signal $X(t-i)$ are taken into account, $X$ is said to Granger-cause $Y$. Time series $X(t)$ and $Y(t)$ can be multivariate, therefore they will be further referred to as $\vec{X}(t)$, $\vec{Y}(t)$.

% A bit more detail
$Y(t)$ is represented as an \emph{autoregressive process}: it is being predicted by a linear combination of its past states and a~Gaussian noise (there is also an~equivalent of GC in the frequency domain, spectral GC~\cite{geweke1982,geweke1984}, but this method will not be covered in this review).
This model is compared to a~model including the past values of $X(t)$:
\begin{equation}
H_0: \vec{Y}(t) = \sum_{i=1}^N \textbf{B}_{yi} \vec{Y}(t-i) + \vec{\sigma}(t) 
\label{eq:gc_eq1}
\end{equation}
\begin{equation}
H_1: \vec{Y}(t) = \sum_{i=1}^N \textbf{B}_{yi} \vec{Y}(t-i) + \sum_{i=1}^N \textbf{B}_{xi} \vec{X}(t-i) + \vec{\sigma}(t)
\label{eq:gc_eq2}
\end{equation}
where $\sigma(t)$ denotes noise (or rather, the~portion of the~signal not explained by the~model). Theoretically, this autoregressive (AR) model can take any order $N$ (which can be optimized using, e.g., Bayesian Information Criterion~\cite{schwarz1978}), but in fMRI research it is usually set to $N = 1$~\cite{seth2015}, i.e. a lag that is equal to the repetition time (TR). 

% Discussing of the Causation List 
% signum (1), strength (2), significance (3)
By fitting the parameters of the~AR model, which include the influence magnitudes $\textbf{B}_{yi}, \textbf{B}_{xi}$, the sign$_{1}$ as well as the strength$_{2}$ of the causal direction can be readily assessed with GC. The significance of the results is evaluated by comparing variance of the~noise obtained from models Eq.~\ref{eq:gc_eq1} and Eq.~\ref{eq:gc_eq2}. This can be achieved either by F-tests or by permutation testing$_{3}$.
% bidirectional (3), immediacy (4), resilience to confounds (5)
Like all the~methods in this chapter, GC does not impose any constraints on the~network architecture and therefore can yield bidirectional connections$_{4}$. As a~multivariate method, GC fits the~whole connectivity structure at once. Therefore, ideally, it indicates the~direct causal connections only$_{5}$, whereas the~indirect connections should be captured only through higher order paths in the~graph revealed in the~GC analysis. However, this is not enforced directly by the method. In fact, in the~original formulation of the problem by Granger, GC between $X$ and $Y$ works based on the assumption that the input of all the~other variables in the~environment potentially influencing $X$ and $Y$ has been removed~\cite{granger1969}. In theory, this would provide resilience to confounds$_{6}$. However, in reality this assumption is most often not valid in fMRI~\cite{moritz-wentrup2014}. In a~result, direct and indirect causality between $X$ and $Y$ are in fact pooled.
% type of inference (6)
In terms of the~inference type, one can look at GC in two ways. On the one hand, GC is a~model-comparison technique, since the~inference procedure is, in principle, based on a comparison between two models expressed by Eqs.~\ref{eq:gc_eq1} and~\ref{eq:gc_eq2}. On the other hand, the~difference between GC and other model comparison techniques lies in the~fact that GC does not optimize any cost function, but uses F-tests or permutation testing instead, and it can therefore also be interpreted as a~method for classic hypothesis testing$_{7}$.
% computational costs (7)
Since the~temporal resolution of fMRI is so low, typically first order AR models with a~time-lag equal to 1 TR are used for the~inference in fMRI. Therefore, there is no need to optimize either the~temporal lag or the~model order, and as such the~computational cost of GC estimation procedure in fMRI is low$_{8}$. The~AR model imposes a mathematical restriction on the size of the network though$_{9}$: the number of regions divided by the number of shifts can never exceed the number of time points (degrees of freedom). 

GC is used in fMRI research in two forms: as mentioned in section~\ref{sec:criteria}, GC can be either applied to the~observed BOLD responses~\cite{zhao2016,regner2016,chen2017}, or to the~BOLD responses \textit{deconvolved} into neuronal time series~\cite{david2008,ryali2011,ryali2016,hutcheson2015,wheelock2014,sathian2013,goodyear2016}. The~purpose of the~deconvolution is to model fMRI data more faithfully. However, estimating the hemodynamic response from the data - a necessity to perform this deconvolution - adds uncertainty as well.

The applicability of GC to fMRI data has been heavily debated~\cite{stokes2017}. Firstly, the~application of GC requires certain additional assumptions such as signal stationarity (stationarity means that the~joint probability distribution in the~signal does not change over time. This also implies that mean, variance and other moments of the~distribution of the~samples in the~signal do not change over time), which does not always hold in fMRI data. Theoretical work by Seth et al.~\cite{seth2013}, and work by Roebroeck et al.~\cite{roebroeck2005}, suggest that despite the~limitations related to slow haemodynamics, GC is still informative about the~directionality of causal links in the~brain~\cite{seth2015}. In the~study by Smith et al.~\cite{smith2011}, several versions of GC implementation were tested. However, all were characterized by a~low sensitivity to correctly connection link detection, low sensitivity to false positives, and low overall accuracy in the~directionality estimation. The~face validity of GC analysis was empirically validated using joint fMRI and MEG recordings~\cite{mill2017}, with the causal links inferred with GC matching the ground truth confirmed by MEG. On the other hand, experimental findings report that GC predominantly identifies major arteries and veins as causal hubs~\cite{webb2013}. This result can be associated with a~regular pulsating behaviour with different phases in the~arteries across the brain. This is a~well-known effect and is even explicitly targeted with physiological noise estimates such as RETROICOR~\cite{Glover2000}.

Another point of concern is the~time lag in fMRI data, which restricts the~possible scope of AR models that can be fit in the~GC procedure. Successful implementations of GC in EEG/MEG research typically involve lags of less than a~hundred milliseconds~\cite{hesse2003}. In contrast, for fMRI the~minimal lag is one full TR, which is typically between $0.7[s]$ and $3.0[s]$ (although new acceleration protocols allow for further reduction of TR). What is more, the HRF may well vary across regions~\cite{handwerker2004, david2008}, revealing spurious causal connections: when the HRF in one region is faster than in another, the temporal precedence of the peak will easily be mistaken for causation. The~estimated directionality can in the worst case, even be reversed, when the region with the slower HRF in fact causes the faster one~\cite{bielczyk2017}. Furthermore, the~BOLD signal might be non-invertible into the neuronal time series~\cite{seth2015}, which can affect GC analysis regardless whether it is performed on the~BOLD time series or the~deconvolved signal.
\subsection{Transfer Entropy}
Transfer Entropy (TE~\cite{schreiber2000}) is another data-driven technique, equivalent to Granger Causality under Gaussian assumptions~\cite{barnett2009}, and asymptotically equivalent to GC for general Markovian
(non-linear, non-Gaussian) systems~\cite{barnett2012}. In other words, TE is a
non-parametric form of GC (or, GC is a~parametric form of TE). It was originally defined for pairwise analysis, and later extended to multivariate analysis~\cite{lizier2008,mute_toolbox}. %In this review, TE is classified as a~`pairwise method' since, due to the binarization of the data, its multivariate version requires time series of length beyond the~range available in fMRI datasets. 
TE is based on the concept of \textit{Shannon entropy}~\cite{shannon1948}. Shannon entropy $H(x)$ quantifies the information contained in a signal of unknown spectral properties as the~amount of uncertainty, or unpredictability. For example, a~binary signal that only gets values of $0$ with a~probability $p$, and values of $1$ with a~probability $1 - p$, is most unpredictable when $p=0.5$. This is because there is always exactly a $50\%$ chance of correctly predicting the next sample. Therefore, being informed about the~next sample in a~binary signal of $p=0.5$ reduces the amount of uncertainty to a~higher extent than being informed about the~next sample in a~binary signal of, say, $p=0.75$. This can be interpreted as a~larger amount of information contained in the~first signal as compared to the~latter. The~formula which quantifies the~information content according to this rule reads as follows:
\begin{equation}
H(X) = - \sum_i P(x_i) \mbox{log}_2P(x_i)
\end{equation}
where $x_i$ are the possible values in the~signal (for the~binarized signal, there are only two possible values: $0$ and $1$).

TE builds up on the~concept of Shannon entropy by extension to \textit{conditional Shannon entropy}: it describes the~amount of uncertainty reduced in future values of $Y$ by knowing the past values of $X$ along with the past values of $Y$:

\begin{equation}
TE_{X \rightarrow Y} = H(Y|Y_{t-\tau}) - H(Y|X_{t-\tau},Y_{t-\tau})
\end{equation}
\noindent where $\tau$ denotes the time lag.  

In theory, TE requires no assumptions about the properties of the data, not even signal stationarity although in most real-world applications, stationarity is required to almost the~same extent as in GC. Certain solutions for TE in non-stationary processes are available though~\cite{wollstadt2014}. TE does need an~a~priori definition of the~causal process, and it may work for both linear and nonlinear interactions between the~nodes.

%-------------------------------------------
% Discussing of the Causation List 
% signum (1), strength (2) and directional (3):
TE can distinguish the~signum of connections$_1$, as the~drop in the~Shannon entropy can be both positive and negative. Furthermore, the~absolute value of the~drop in the Shannon entropy can provide a~measure of the~connection strength$_2$. TE can also distinguish bidirectional connections, as in this case, both $TE_{X\rightarrow Y}$ and $TE_{Y\rightarrow X}$ will be nonzero$_4$. 
% significance testing (3):
In TE, significance testing by means of permutation testing is advised~\cite{vicente2011}$_3$.
% immediacy (4) and resilience to confounds (5):
Immediacy and resilience to confounds in TE depends on the implementation to a~large extent: using a simple Pearson's correlation to compute functional connectivity increases the amount of spurious (indirect) connections, whereas partial correlation is meant to pick up on direct connections only$_{5,6}$. 
% type of inference (6) and computational cost (7):
The~inference in TE is performed through classical hypothesis testing$_7$ and is highly cost-efficient$_8$.  
As in GC, the maximum number of regions in the~network divided by the number of shifts can never exceed the number of time points (degrees of freedom)$_9$. 
%------------------------------------------

TE is a~straightforward and computationally cheap method~\cite{vicente2011}. However, it struggled when applied to synthetic fMRI benchmark datasets~\cite{smith2011}. One reason for this could be the~time lag embedded in the~inference procedure, which is an~obstacle to TE in fMRI research for the~same reasons as for GC: it requires at least one full TR. TE is nevertheless gaining interest in the~field of fMRI~\cite{sharaev2016,lizier2011,ostwald2011,chai2009,mute_toolbox}.
\subsection{Structural Equation Modeling}~\label{sec:sem}
Structural Equation Modeling (SEM,~\cite{mcintosh1994}) is a~simplified version of Granger Causality. This method was originally applied to a~few disciplines: economics, psychology and genetics~\cite{wright1920}, and was only recently adapted for fMRI research~\cite{mcintosh1994}. SEM can be considered a~predecessor to Dynamic Causal Modelling~\cite{friston2003}. SEM is used to study effective connectivity in cognitive paradigms, e.g., on motor coordination~\cite{kiyama2014,zhuang2005}, as well as in search for biomarkers of psychiatric disorders~\cite{schlosser2003,carballedo2011}. It was also used for investigating heritability of large scale, resting state connectivity patterns~\cite{carballedo2011}.

The idea is to express every ROI time series in a~network by a~\textit{linear combination} of all the~time series (with the~addition of noise), which implies no time lag in the~communication. These signals are combined in a~mixing matrix~$\textbf{B}$:
\begin{equation}
\vec{X}(t) = \textbf{B} \vec{X}(t) + \vec{\sigma}(t)
\label{eq:linear}
\end{equation}
\noindent where $\vec{\sigma}$ denotes the noise, and the~assumption is that each univariate component $X_i(t)$ is a~mixture of the~remaining components $X_j(t)$, $j \neq i$. This is a~simple multivariate regression equation. The most common form of fitting this model is a~search for the regression coefficients that corresponds to the~maximum likelihood (ML) solution: a~set of model parameters $\textbf{B}$ that gives the highest probability of the observed data~\cite{mcintosh1994,anderson1988}. Assuming that variables $X_i$ are normally distributed, the~ML function can be computed and optimized. This function is dependent on the~observed covariance between variables, as well as a~concept of a~so-called \textit{implied} covariance, for the details, see~\cite{bollen1989}, and for a~practical example of SEM inference, see~\cite{ferron2007}. Further, under the assumption of normality of the noise, there is a~closed-form solution to this problem which gives the ML solution for parameters $\textbf{B}$, known as Ordinary Least Squares (OLS) approximation~\cite{hayashi2000,bentler1985}. %In OLS, we are looking for $\textbf{B}$ parameters by minimizing the term 
%\begin{equation}
%\|\vec{X}-\textbf{B}\vec{X}\|^2
%\end{equation}

In SEM applications to fMRI datasets, it is a~common practice to establish the~presence of connections with use of anatomical information derived, e.g., from Diffusion Tensor Imaging~\cite{protzner2006}. In that case, SEM inference focuses on estimating the~strength of causal effects and not on identifying the~causal structure.

%-------------------------------------------
% Discussing of the Causation List 
% signum (1) and directional (3):
SEM does not constrain the~weight of connections, therefore it can retrieve both excitatory and inhibitory connections$_1$ as well as bidirectional connections$_4$. 
% strength (2):
The~connection coefficients $\textbf{B}_{ij}$ can take any rational numbers and as such, they can reflect the~strength of the~connections$_2$. 
% significance testing (3):
Since OLS gives a~point estimate for $\beta$, it does not provide a measure of confidence that would determine whether the~obtained $\beta$ is significantly different from zero. This issue can be overcome in multiple ways. First, one can perform parametric tests, e.g., a~t-test. Second, one can obtain confidence intervals through nonparametric permutation testing (generate a~null distribution of $\textbf{B}$ values by the repeated shuffling of node labels across subjects and creating surrogate subjects). Third, one can perform causal inference through model comparison: various models are fitted one by one, and the variance of the~residual noise resulting from different model fits is compared, using either an~F-test, or goodness of fit (GFI,~\cite{zhuang2005}). Highly optimized software packages such as LiSREL~\cite{joreskog1972} allow for an~exploratory analysis with SEM by comparing millions of models against each other~\cite{james2009}. Lastly, one can fit the $\textbf{B}$ matrix with new methods including regularization that enforces sparsity of the~solution~\cite{jacobucci2016}, and therefore eliminates weak and noise-induced connections from the~connectivity matrix$_3$.
% immediacy (4)
As with GC, SEM was designed to reflect direct connections$_5$: if regions $X_i$ and $X_j$ are connected only through a~polysynaptic causal web, $\textbf{B}_{ij}$ should come out as zero, and the polysynaptic connection should be retrievable from the~path analysis.
% resilience to confounds (5)
Again similar to GC, SEM is resilient to confounds only under the~assumption that the model represents an~isolated system, and all the relevant variables present in the~environment are taken into account$_{6}$. Moreover, in order to obtain the maximum likelihood solution for $\textbf{B}$ parameters, one needs to make a~range of assumptions on the~properties of the~noise in the~network. Typically, a~Gaussian white noise is assumed, although background noise in the~brain is most probably scale-free~\cite{he2014}. 
% type of inference (6) and computational costs (7)
Inference can be performed either through the classical hypothesis testing (as the~computationally cheap version) or through model comparison (as the~ computationally heavier version)$_{7,8}$.
%------------------------------------------ 

In summary, SEM is a~straightforward approach: it simplifies the~causal inference by reducing the~complex network with a~low-pass filter at the~output to a~very simple linear system, but this simplicity comes at the cost of a number of assumptions. In the~first decade of fMRI research, SEM was often a~method of choice~\cite{zhuang2008,schlosser2008} however recently, using Dynamic Causal Modeling has become more popular in the~field. One recently published approach in this domain, by Schwab et al.~\cite{schwab2018}, extends linear models by introducing time-varying connectivity coefficients, which allows for tracking the dynamics of causal interactions over time. In this approach, linear regression is applied to each node in the network separately (in order to find causal influence of all the remaining nodes in the network on that node). The whole graph is then composed from node-specific DAGs node by node, and that compound graph can be cyclic. 
\subsection{Dynamic Causal Modeling}\label{sec:dcm}
% Introduce DCM
Both the aforementioned network-wise methods were developed in other disciplines, and only later applied to fMRI data. Yet, using prior knowledge about the properties of fMRI datasets can prove useful when searching for causal interactions. Dynamic Causal Modeling (DCM~\cite{friston2003}) is a~hypothesis testing tool which uses state space equations reflecting the~structure of fMRI datasets. This technique was also implemented for other neural recording methods: EEG and MEG~\cite{kiebel2008}. DCM is well received within the neuroimaging community (the original article by Friston et al.~\cite{friston2003} gained over 2,700 citations at the time of submitting this manuscript).

% This is what it is
In this work, we describe the~original work by Friston et al.~\cite{friston2013} because, despite multiple recent developments~\cite{kiebel2007, stephan2007, marreiros2008, stephan2008, li2011, daunizeau2012, seghier2013, friston2014, havlicek2015, frassle2016, razi2016, prando2017, frassle2017}, it remains the~most popular version of DCM in the fMRI community. The idea of DCM is as follows. First, one needs to build a~generative forward model (Fig.~2). This model has two levels of description: the neuronal level (Fig.~2, (iii)), and the haemodynamic level (Fig.~2, (v)). Both of these levels contain parameters which are not directly recorded in the~experiment and need to be inferred from the data. This model reflects scientific evidence on how the~BOLD response is generated from neuronal activity.

% More detailed explanation of previous paragraph
At the~neuronal level of the~DCM generative model, simple interactions between brain areas are posited, either
bilinear~\cite{friston2003} or nonlinear~\cite{stephan2008}. In the~simplest, bilinear version of the~model, the~biliear state equation reads:

\begin{equation}
\dot{z} = (\textbf{A} + \sum_{j}u_j\textbf{B}^j)z + \textbf{C}u
\label{eq:DCM_state_equation}
\end{equation}

\begin{figure}[h]
\begin{framed}
\centering
\includegraphics[width=0.85\textwidth]{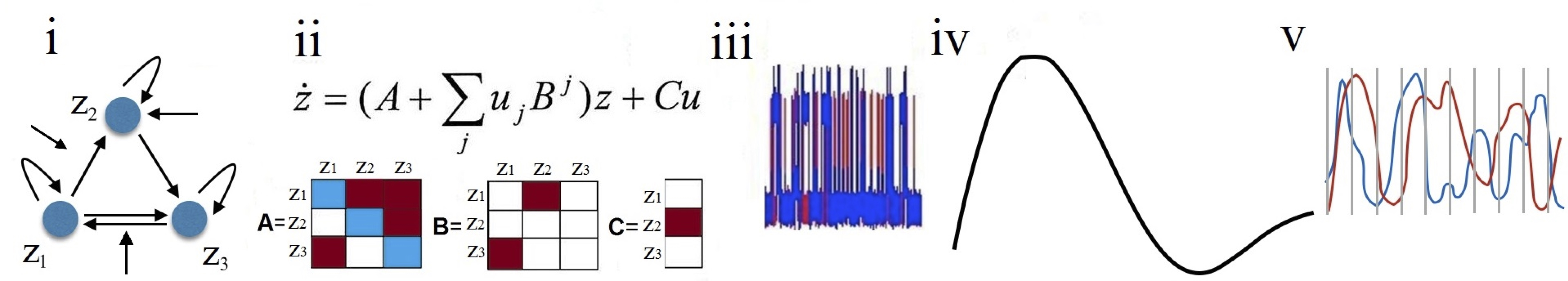}
\end{framed}
\caption{The full pipeline for the DCM forward model. The~model
involves three node network stimulated during the~cognitive experiment~\textbf{(i)}. The~parameter set describing the~dynamics in this network includes a~fixed connectivity matrix (A), modulatory connections (B), and inputs to the~nodes (C)~\textbf{(ii)}. In the~equation describing the~fast neuronal dynamics, $z$ denotes the dynamics in the~nodes, and $u$ is an~experiment-related input. Red: excitatory connections. Blue: inhibitory connections. The~dynamics in this network can be described with use of ordinary differential equations. The~outcome is the~fast neuronal dynamics~\textbf{(iii)}. The~neuronal time series is then convolved with the haemodynamic response function~\textbf{(iv)} in order to obtain the~BOLD response~\textbf{(v)}, which may be then subsampled (vertical bars). This is the~original, bilinear implementation of DCM~\cite{friston2003}. Now, more complex versions of DCM with additional features are available, such as spectral DCM~\cite{friston2014}, stochastic DCM~\cite{daunizeau2012}, nonlinear DCM~\cite{stephan2008}, two-state DCM~\cite{marreiros2008}, large DCMs~\cite{seghier2013,frassle2016} etc.}
\label{fig:fig2}
\end{figure}

\noindent where $z$ denotes the~dynamics in the~nodes of the~network, $u$ denotes the~experimental inputs, $A$ denotes the~connectivity matrix characterizing causal interactions between the~nodes of the~network, $B$ denotes the~modulatory influence of experimental inputs on the~connections within the~network, and $C$ denotes the~experimental inputs to the~nodes of the~network (Fig.~2). The~haemodynamic level is more complex and follows the biologically informed Balloon-Windkessel model~\cite{buxton1998}, for details please see~\cite{friston2003}. The~Balloon-Windkessel model~\cite{buxton1998} describes the BOLD signal observed in fMRI experiments as a~function of neuronal activity but also region-specific and subject-specific physiological features such as the time constant of signal decay, the~rate of flow-dependent elimination, and the~haemodynamic transit time or resting oxygen fraction. This is a~weakly non-linear model with free parameters estimated for each brain region. These parameters determine the~shape of the~hemodynamic response (Fig.~2, (iv)), which typically peaks at $4-6[s]$ after the neuronal activity takes place, to match the~lagged oxygen consumption in the~neuronal tissue mentioned in Section~\ref{sec:limitations_hemod}. The~Balloon-Windkessel model is being iteratively updated based on new experimental findings, for instance to mimic adaptive decreases to sustained inputs during stimulation or the~post-stimulus undershoot~\cite{havlicek2015}.

% More about the estimation procedure
In this paper, the~deterministic, bilinear single state per region DCM will be described~\cite{friston2003}. The~DCM procedure starts with defining hypotheses based on observed activations, which involves defining which regions are included in the network (usually on the basis of activations found through the General Linear Model~\cite{friston_book}) and then defining a~model space based on the~research hypotheses. In the latter model selection phase, a~range of literature-informed connectivity patterns and inputs in the~networks (referred to as 'models') are posited (Fig.~2, (i)). The definition of a~model space is the~key to the~DCM analysis. The~models should be considered carefully in the~light of the~existing literature. The~model space represents the formulation of a~\textit{prior} over models, therefore, it should always be constructed \textit{prior} to the~DCM analysis. Subsequently, for every model one needs to set priors on the parameters of interest: connectivity strengths and input weights in the model (Fig.~2, (ii)) and the~haemodynamic parameters. The~priors for haemodynamic parameters are experimentally informed Gaussian distributions~\cite{friston2003}. The priors for connectivity strengths are Gaussian probability distributions centered at zero (which is often referred to as \textit{conservative shrinkage} priors). The user usually does not need to specify the priors, as they are already implemented in the~DCM algorithms. 

Next, an~iterative procedure is used to find the model evidence by maximizing a~cost function, a~so-called \textit{negative free energy}~\cite{friston2007}. Negative free energy is a~particular cost function which gives a~trade-off between model accuracy and complexity (which accounts for correlations between parameters, and for moving away from the prior distributions). During the~iterative procedure, the prior probability distributions gradually shift their mean and standard deviation, and converge towards the final posterior distributions. Negative free energy is a more sophisticated approximation of the model evidence when compared to methods such as Akaike's Information Criterion (AIC,~\cite{akaike1998information}) or Bayesian Information Criterion (BIC,~\cite{schwarz1978}); AIC and BIC simply count the number of free parameters (thereby assuming that all parameters are independent), while negative free energy also takes the covariance of the parameters into account~\cite{penny2012}.

% -------------------------------------
% Discussing of the Causation List 
% excitation/inhibition:
In DCM, causality is modeled as a~set of upregulating or downregulating connections between nodes. During the~inference procedure, conservative shrinkage priors can shift towards both positive and negative values, which can be interpreted as effective excitation or effective inhibition (except for self connections, which are always only negative (this~self inhibition is mathematically motivated: the~system characterizing the~fast dynamics of the neuronal network must be stable, and this requires the diagonal terms of the adjacency matrix $A$ to be negative), Fig.~2, (ii), connections denoted in blue)$_1$. 
During the~inference procedure, the~neural and hemodynamic parameters of all models postulated for model comparison are optimized$_2$. The posterior probability distributions determine significance of all the~parameters$_3$. The models can contain both uni- and bidirectional connections~\cite{vaudano2013, buijink2015}$_4$. The~estimated model evidence can then be compared$_7$. As such, the~original DCM~\cite{friston2003} is a~hypothesis-testing tool working only through model comparison. However, now, a~linear version of DCM dedicated to exploratory research in large networks is also available~\cite{frassle2016}. Testing the immediacy$_5$ and resilience to confounds$_6$ in DCM is possible through creating separate models and comparing their evidence. For instance, one can compare the~evidence for $X \rightarrow Y$ with evidence for $X \rightarrow Z \rightarrow Y$ in order to test whether or not the connection $X \rightarrow Y$ is direct or rather mediated by another region $Z$. Note that this strategy requires an~explicit specification of the~alternative models and it cannot take hidden causes into consideration (in this work, we refer to the~original DCM implementation~\cite{friston2003}, but there are also implementations of DCM involving estimation of time-varying hidden states, such as~\cite{daunizeau2009}). 
However, including extra regions in order to increase resilience to confounds is not necessarily a~good idea. Considering the potentially large number of fitted parameters per region (the~minimum number of nodes per region is two hemodynamic parameters and one input/output to connect to the rest of network), this may result in a~combinatorial explosion. Also, models with different nodes are not comparable in DCM for fMRI~\cite{friston2003}. Extending the models by adding additional nodes not only increases the computation time considerably$_8$. The~original DCM~\cite{friston2003} is therefore restricted to small networks of a~few nodes$_{9}$ (as mentioned previously, today, large DCMs dedicated to exploratory research in large networks are also available~\cite{seghier2013,frassle2016}).

% ------------------------------------

% model specification:
The proper application of DCM needs a~substantial amount of expertise~\cite{stephan2010,daunizeau2011}. Even though ROIs can be defined in a~data-driven fashion (through a~preliminary classical General Linear Model analysis~\cite{friston1995}), the~model space definition requires prior knowledge of the research problem~\cite{kahan2013}. In principle, the~model space should reflect prior knowledge about possible causal connections between the~nodes in the~network. If a~paradigm developed for the~fMRI study is novel, there might be no reference study that can be used to build the~model space. In that case, using \textit{family-wise} DCM modeling can be helpful~\cite{penny2010}. Family-wise models group large families of models defined on the~same set of nodes, in order to test a~particular hypothesis. For instance, one can explore a~three node network with nodes $X$, $Y$, $Z$ and compare the~joint evidence behind all the~possible models that contain connection $X \rightarrow Y$ with the joint evidence behind all the~possible models that contain connection $Y \rightarrow X$ (Fig.~2, (i)). Another solution that allows for constraining a~large model space is Bayesian model averaging (BMA,~\cite{hoeting1999,stephan2010}) which explores the~entire model space and returns \textit{average} value for each model parameter, weighted by the~posterior probability for each model. Finally, one can perform a Bayesian model reduction~\cite{friston2016}, in which the considered models are reduced versions of a~full (or 'parent') model. This is possible when the priors can be reduced, e.g. when a~prior distribution of a~parameter in a~parent model is set to a~mean and variance of zero. 

The~number of possible models explodes with the~size of the~network. In order to extend the~scope of application of the~DCM analysis to larger networks, recently the~new, large-scale DCM framework for resting state fMRI has been proposed~\cite{razi2017}. This framework uses the~new, spectral DCM~\cite{friston2014} designed for resting state fMRI and which is able to handle dozens of nodes in the~network. Spectral DCM is then combined with functional connectivity priors in order to estimate the~effective connectivity in the~large-scale resting state networks.

% Model comparison issues:
There are a few points that need particular attention when interpreting the results of the DCM analysis. Firstly, in case the data quality is poor, evidence for one model over another will not be conclusive. In the worst case, it could give a preference to the simplest model (i.e. the model with the fewest free parameters). In that case, simpler models will be preferred over more complex ones regardless of the low quality of fit. It is important, therefore, to include a 'null model' in a DCM analysis, with all interesting parameters fixed at zero. This can then act as a baseline, against which models of interest can be compared~\cite{penny2012}. %Therefore, it is recommended to iterate the process of choosing the~model space until the~winning model is in the middle of the range in terms of complexity.
% ---------------------------------------------------------------------------

% Model comparison issue, specific to DCM
Second, the winning model might contain parameters with a~high probability of being equal to zero. To illustrate this, let us consider causal inference in a~single subject (also referred to as \textit{first level} analysis). Let us assume that we chose a~correct set of priors (model space). The Variational Bayes procedure then returns a posterior probability distribution for every estimated connectivity strength. This distribution gives a measure of probability for the associated causal link to be larger than zero. Some parameters may turn out to have high probability of being equal to zero in the light of this posterior distribution. This may be due to the fact that the winning model is correct, but some of the underlying causal links are weak and therefore hard to confirm by the VB procedure. Also, DCM requires data of high quality; when the SNR is insufficient, it is possible that the winning model would explain a small portion of the variance in the data. In that case, getting insignificant parameters in the winning model is likely. Therefore, it is advisable to check the amount of variance explained by the winning model at the end of the DCM analysis.

% Variational Bayes vs MCMC
The~most popular implementation of the~DCM estimation procedure is based on Variational Bayes (VB,~\cite{bishop2006}) which is a~deterministic algorithm. Recently, also Markov-Chain Monte Carlo (MCMC,~\cite{bishop2006,sengupta2015} was implemented for DCM. When applied to a~unimodal free energy landscape, these two algorithms will both identify the~global maximum. MCMC will be slower than VB as it is stochastic and therefore computationally costly. However, free energy landscape for multiple-node networks is most often multimodal and complex. In such case, VB - as a~local optimization algorithm - might settle on a~\textit{local} maximum. MCMC on the other hand, is guaranteed to converge to the true posterior densities - and thus the~\textit{global} maximum (given an infinite number of samples).  

% DCM validity and reliability, still developing:
DCM was tailored for fMRI and, unlike other methods, it explicitly models the haemodynamic response in the~brain. The~technique tends to return highly reproducible results, and is therefore statistically reliable~\cite{schuyler2010, rowe2010, bernal2013, tak2018}. Recent longitudinal study on spectral DCM in resting state revealed systematic and reliable patterns of hemispheric asymmetry~\cite{almgren2018}. DCM also yielded high test-retest reliability in an~fMRI motor task study~\cite{frassle2015}, in a~face perception study~\cite{frassle2016a}, in facial emotion perception study~\cite{schuyler2010} and in a~finger-tapping task in a~group of subjects suffering from Parkinson's disease~\cite{rowe2010}. It has also been demonstrated most reliable when directly compared to GC and SEM~\cite{penny2004}. Furthermore, the~DCM procedure can provide complimentary information to GC~\cite{friston2013}: GC models dependency among observed BOLD responses, whereas DCM models coupling among the~hidden states generating observations. GC seems to be equally effective as DCM in certain circumstances, such as when the~haemodynamic response function (HRF) is deconvolved from the~data~\cite{david2008,ryali2011,ryali2016,wang2016a}. Importantly, the~face validity of DCM was examined on experimental datasets coming from interventional study with use of rat model of epilepsy~\cite{david2008,papadopoulou2015}.

On the~other hand, proper use of DCM requires knowledge on the~biology and on the~inference procedure. DCM also has limitations in terms of the size of the possible models. Modeling a large network may run into problems with identifiability - there will be many possible combinations of parameter settings which could give rise to the same or similar model evidence. In other words, strong covariance between parameters will preclude confident estimates of the strength of each connection. One possible remedy for this, in the context of large scale networks, is to impose appropriate prior constrains on the connections - for example, using priors based on functional connectivity as priors~\cite{razi2017}. Large networks may also give rise to comparisons of large number of different models with varying combinations of connections. To reduce the possibility of overfitting at the level of model comparison - i.e. finding a model which is appropriate for one subject or group of subjects' data, but not for others - it can be useful to group the models into a small number of families~\cite{penny2010} based on pre-defined hypotheses. More information on the~limitations of DCM can be found in work by Daunizeau et al.~\cite{daunizeau2011} (a critical note on limitations of DCM in terms of network size can also be found in~\cite{lohmann2012}, see also a~response to this article,~\cite{friston2013a,breakspear2013}). 

However, in order to extend the~scope of application of the~DCM analysis to larger networks, recently two approaches have been proposed. Firstly, a~new, large-scale DCM framework for resting state fMRI has been proposed~\cite{razi2017}. This framework uses the~new, spectral DCM~\cite{friston2014} designed for resting state fMRI and which is able to handle dozens of nodes in the~network. Spectral DCM is then combined with functional connectivity priors in order to estimate the~effective connectivity in the~large-scale resting state networks. Secondly, a~new approach by Fraessle et al.~\cite{frassle2018} imposes sparsity constraints on the~variational Bayesian framework for task fMRI, which enables for causal inference on the~whole-brain network level.

DCM was further developed into multiple procedures including more sophisticated generative models than the original model discussed here. The~field of DCM research in fMRI is still growing~\cite{friston2017}. The~DCM generative model is continuously being updated, in terms of the~structure of the~forward model (\cite{havlicek2015}, the~estimation procedure\cite{sengupta2015}), and the~scope of the~possible applications \cite{friston2017}.

% [b] network level + DAG
\section{Hierarchical network-wise models}
The second group of methods involves hierarchical network-wise models: Linear Non-Gaussian Acyclic Models (LiNGAM,~\cite{shimizu2006}) and Bayesian Nets~\cite{frey2005}. Similarly as network-wise methods reviewed in the~previous chapter, these methods are also multivariate but with one additional constraint: the~network can only include \textit{feed forward} projections (and therefore, no closed cycles). Consequently, the resulting models have a hierarchical structure with feed forward distribution of information through the network. 
\subsection{LiNGAM}\label{sec:lingam}
The Linear Non-Gaussian Acyclic Model (LiNGAM, ~\cite{shimizu2006}) is an~example of a~data driven approach working under the~assumption of acyclicity~\cite{thulasiraman1992}. The model itself is simple: every time course within an~ROI $X_i(t)$ is considered to be a~linear combination of all other signals with no time lag:
\begin{equation}
\vec{X}(t) = \textbf{B} \vec{X}(t) + \vec{\sigma}(t)
\label{eq:ica_model}
\end{equation}
in which $\textbf{B}$ denotes a~matrix containing the connectivity weights, and $\vec{\sigma}$ denotes noise. The model is in principle the same as in SEM (Section~\ref{sec:sem}), but the~difference lies in the~inference procedure: whereas in SEM, inference is based on minimizing the \textit{variance} of the~residual noise under the assumption of independence and Gaussianity, LiNGAM finds connections based on the~\textit{dependence} between residual noise components $\vec{\sigma}(t)$ and regressors $\vec{X}(t)$. 

% 'LiNGAM story':
The~rationale of this method is as follows (Fig. 3). Let us assume that the~network is noisy, and every time series within the network is associated with a~background noise uncorrelated with the~signal in that node. An~example of such a~mixture of signal with noise is given in Fig.~3A. Then, let us assume that $\hat{X}(t)$ - which is a~mixture of signal $X(t)$ and noise $\sigma_X(t)$ - causes $Y(t)$. Then, as it cannot distinguish between the~signal and the~noise, $Y$ becomes a~function of both these~components. $Y(t)$ is also associated with noise $\sigma_Y(t)$, however, as there is no causal link $Y \rightarrow X$, $X(t)$ is not dependent on the~noise component $\sigma_Y(t)$. Therefore, if $Y$ depends on the~$\sigma_X(t)$ component, but $X$ does not depend on the~$\sigma_Y(t)$ component, one can infer projection $X \rightarrow Y$.

\begin{figure}[h]
\begin{framed}
\centering
\includegraphics[width=0.85\textwidth]{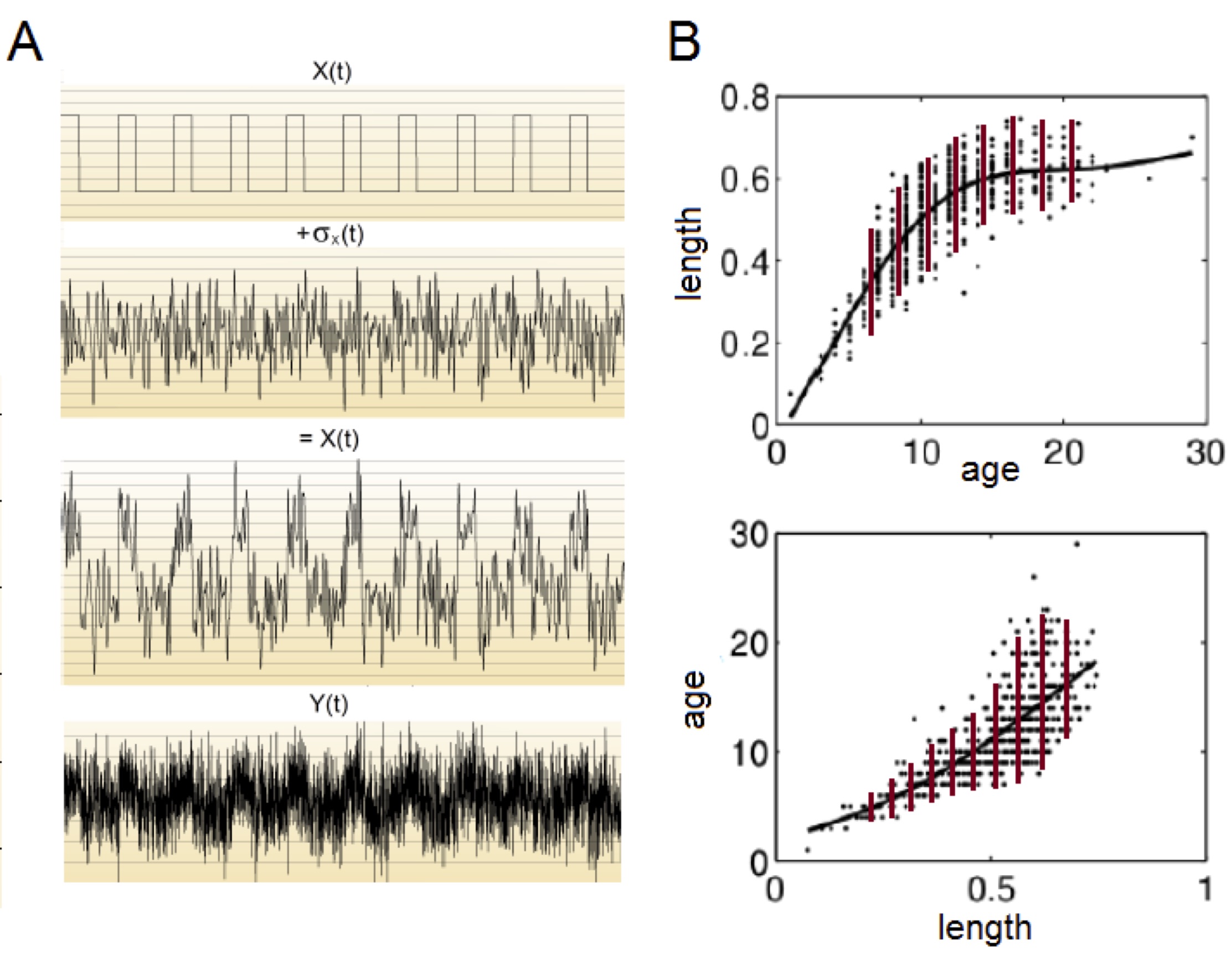}
\end{framed}
\caption{LiNGAM. \textbf{A}: The~noisy time series $\hat{X}(t)$ consists of signal $X(t)$ and noise $\sigma_X(t)$. $Y(t)$ thus becomes a~function of both the~signal and the~noise in $\hat{X}(t)$. \textbf{B}: Causal inference through the analysis of the~noise residuals (figure reprinted from \url{http://videolectures.net/bbci2014_grosse_wentrup_causal_inference/}). The~causal link from age to length in a~population of fish can be inferred from the properties of the residual noise in the~system. the~relationship between age and length in a~fish. If fish length is expressed in a~function of fish age (upper panel), the residual noise in the~dependent variable (length) is uncorrelated with the independent variable (age): the noise variance is constant over a~large range of fish age (red bars). On the~contrary, once the variables are flipped and fish age becomes a~function of fish length (lower panel), the~noise variance becomes dependent on the independent variable (length) as it is small for small values of fish length and large for the large values of fish length (red bars).}
\label{fig:fig3}
\end{figure}

An~example of such a~simple, directed causal relationship between two variables is demonstrated in Fig.~3B: the~relationship between age and length in a~fish. If fish length is expressed in a~function of fish age (upper panel), the residual noise in the~dependent variable (length) is uncorrelated with the independent variable (age). Therefore, the noise variance is constant over a~large range of fish age. On the~contrary, once the variables are flipped and fish age becomes a~function of fish length (lower panel), the~noise variance becomes dependent on the independent variable (length) as it is small for small values of fish length and large for the large values of fish length. Therefore, the~first causal model (fish age influencing fish length) is correct. 

In applications to causal research in fMRI, the~LiNGAM inference procedure is often accompanied by an Independent Component Analysis (ICA,~\cite{hyvarinen2000}) as follows. The~connectivity matrix $\textbf{B}$ in Eq.~\ref{eq:ica_model} describes how signals in the~network mix together. By convention, not $\textbf{B}$ itself but a~transformation of $\textbf{B}$ into 
\begin{equation}
\textbf{A} = (1-\textbf{B})^{-1} 
\label{eq:Amixing}
\end{equation}
is used as a~\textit{mixing matrix} in the~LiNGAM inference procedure. By using this mixing matrix $\textbf{A}$, one can look at Eq.~\ref{eq:ica_model} in a~different way:
\begin{equation}
\vec{X} = \textbf{A}\vec{\sigma}
\label{eq:ica_mixing}
\end{equation}

Now, the~BOLD time course in the~network $\vec{X}(t)$ can be represented as a~mixture of independent \textit{sources of noise} $\vec{\sigma}(t)$. This is the~well known \textit{cocktail party problem} and it was originally described in acoustics~\cite{bronkhorst2000}: in a~crowded room, a~human ear registers a~linear combination of the~noises coming from multiple sources. In order to decode the~components of this cacophony, the~brain needs to perform a~\textit{blind source separation}~\cite{comon2010}: to decompose the~incoming sound into a~linear mixture of independent sources of sounds. In the~LiNGAM procedure, Independent Component Analysis (ICA,~\cite{hyvarinen2000}) is used to approach this issue. ICA assumes that the~noise components $\vec{\sigma}$ are independent and have a~\textit{non-Gaussian distribution}, and finds these components as well as the~mixing matrix $\textbf{A}$ through dimensionality reduction with Principal Component Analysis~\cite{jolliffe2002,schlens2014b}. From this mixing matrix, one can in turn estimate the~desired adjacency matrix $\textbf{B}$ with use of Eq.~\ref{eq:Amixing}.

Since the~entries $\textbf{B}_{ij}$ of the connectivity matrix $\textbf{B}$ can take any value, LiNGAM can in principle retrieve both excitatory and inhibitory connectivity$_{1}$ of any strength$_{2}$. 
% significance testing:
The~author of LiNGAM recommends~\cite{shimizu2014} performing significance testing through either bootstrapping~\cite{hyvarinen2010, komatsu2010, thamvitayakul2012} or permutation testing~\cite{hyvarinen2013} $_{3}$.
% Directionality:
However, LiNGAM makes the~assumption of acyclicity, therefore only unidirectional connections can be picked up$_{4}$.
% Immediacy:
Moreover, the connectivity matrix revealed with the use of LiNGAM is meant to pick up on direct connections$_{5}$.
% Resilience to confounds:
The original formulation of LiNGAM assumes no latent confounds~\cite{shimizu2006}, but the model can be extended to a~framework that can capture the causal links even in the~presence of (unknown) hidden confounds~\cite{hoyer2008, chen2013}$_{6}$. 
% Type of inference:
LiNGAM-ICA's causal inference consists of ICA and a~simple machine learning algorithm, and, as such, it is a~fully data driven strategy that does not involve model comparison$_{7}$. Confidence intervals for the~connections $\textbf{B}$ can be found through permutation testing.  
% Computational cost: 
ICA itself can be computationally costly and its computational stability cannot be guaranteed (the procedure that searches for independent sources of noise can get stuck in a~local minimum). Therefore, the~computational cost in LiNGAM can vary depending on the~dataset$_{8}$. %Furthermore, since the dimensionality reduction with Principal Component Analysis is performed in the first step of ICA, it might reduce the dimensionality of the network by finding fewer independent components than the number of provided variables, which is an~obstacle in practical applications.
This also sets a limit on the potential size of the causal network. When the number of connections approaches the number of time points (degrees of freedom), the fitting procedure will become increasingly unstable as it will be overfitting the data$_{9}$.
% ---------------------------------------------------------------

When tested on synthetic fMRI benchmark datasets~\cite{smith2011}, LiNGAM-ICA performs relatively good , but is more sensitive to confounders than several other methods discussed in this paper, such as Patel's tau or GC. However, as LiNGAM performs particularly well for datasets containing a large number of samples, the~authors suggested that a~\textit{group} analysis could resolve the~sensitivity problem in LiNGAM. The~concept was then picked up and developed by at least two groups. Firstly, Ramsey et al.~\cite{ramsey2011} proposed \textit{LiNG Orientation, Fixed Structure} technique (LOFS). The~method is inspired by LiNGAM, and uses the~fact that, within one graph equivalence class, the~correct causal model should return conditional probability distributions that are maximally non-Gaussian. LOFS was tested on the~synthetic benchmark datasets, where it achieved performance very close to $100\%$. Secondly, Xu et al. published a~pooling-LiNGAM technique~\cite{xu2014}, which is a~classic LiNGAM-ICA applied to the~surrogate datasets. Validation on synthetic datasets revealed that both False Positive (FP) and False Negative (FN) rates decrease exponentially along with the~length of the~(surrogate) time series, however, combining time series of as long as 5,000 samples is necessary for this method to give both FP and FN as a~reasonable level of 5\%.

Despite the~promising results obtained in the~synthetic datasets, LiNGAM is still rarely applied to causal research in fMRI to date.
\subsection{Bayesian nets}
The use of the~LiNGAM inference procedures assumes a~linear mixing of signals underlying a~causal interaction. Model-free methods do not make this assumption: the~bare fact that one is likely to observe $Y$ given the presence of $X$ can indicate that the~causal link $X \rightarrow Y$ exists (Fig. 4). Let as assume the~simplest example: causal inference for two binary signals $X(t)$, $Y(t)$. In a~binary signal, only two values are possible: $1$ and $0$; $1$ can be interpreted as an~'event' while $0$ - as 'no event'. Then, if in signal $Y(t)$, events occur in 80\% of the~cases when events in signal $X(t)$ occur (Fig.~4A), but the opposite is not true, the~causal link $X \rightarrow Y$ is likely. Computing the~odds of events given the~events in the~other signal, is sufficient to establish causality. In a~model-based approach on the~other hand, a~\textit{model} is fitted to the~data, in order to establish the~precise form of the~influence of the~independent variable $X$ on the~dependent variable $Y$. 

Note that both model-based and model-free approaches contain a~measure of uncertainty, but this uncertainty is computed differently. In model-based approaches,~p-values associated with the~fitted model are a~measure of confidence that the modelled causal link is a~true positive (Fig.~4A, left panel). In contrast, in model-free approaches this confidence is quantified directly by quantifying causal relationships in terms of conditional probabilities (Fig.~4A, right panel). In practice, since the~BOLD response - unlike the~aforementioned example of binary signals - takes continuous values, estimating conditional probabilities is based on the~basis of the~\textit{joint distribution} of the~variables $X$ and $Y$ (Fig.~4B). Conditional probability $P(Y|X)$ becomes a~distribution of $Y$ when $X$ takes a~given value. Bayesian Networks (BNs~\cite{frey2005}) are based on such a~model-free approach (Fig.~4C).

The~causal inference in BNs is based on the~concept of \textit{conditional independency} (a.k.a. Causal Markov Condition~\cite{hausman1999}). Suppose that there are two events that could independently cause the~grass to get wet: either a sprinkler, or rain. When one only observes the~grass being wet, the direct cause for this event is unknown. However, once rain is observed, it becomes less likely that the sprinkler was used. Therefore, one can say that the variables $X_1$ (sprinkler) and $X_2$ (rain) are conditionally dependent given variable $X_3$ (wet grass), because $X_1$, $X_2$ become dependent on each other after information about $X_3$ is provided. In BNs, the~assumption of conditional dependency in the~network is used to compute the joint probability of a given model - i.e. the~model evidence (once variables $X_i$ are conditionally dependent on $X_j$, the joint distribution $P(X_i,X_j)$ factorizes into a~product of probabilities $P(X_j)P(X_i|X_j)$). 

Implementing a~probabilistic BN requires defining a model: choosing a~graph of `parents' who send information to their~`children'. For instance, in Fig.~4C, (i), the~node $X_1$ is a~parent of nodes $X_4$ and $X_5$, and the~node $X_4$ is a~child of nodes $X_1$, $X_2$ and $X_3$. The joint probability of the model can then be computed as the product of all marginal probabilities of the parents and conditional probabilities of the children given the parents. \textit{Marginal probability} $P(X_j)$ is the total probability that the variable of interest $X_j$ occurs while disregarding the values of all the other variables in the system. For instance, in Fig.~4C, (i), $P(X_1)$ means a~marginal probability of $X_1$ happening in this experiment. \textit{Conditional probability} $P(X_i|X_j)$ is the probability of a~given variable ($X_i$) occurring given that another variable has occurred ($X_j$). For instance, in Fig.~4C, (i), $P(X_5|X_1, X_3)$ means a~conditional probability of $X_5$ given its parents $X_1$ and $X_3$.

\begin{figure}[H]
\begin{framed}
\centering
\includegraphics[width=0.85\textwidth]{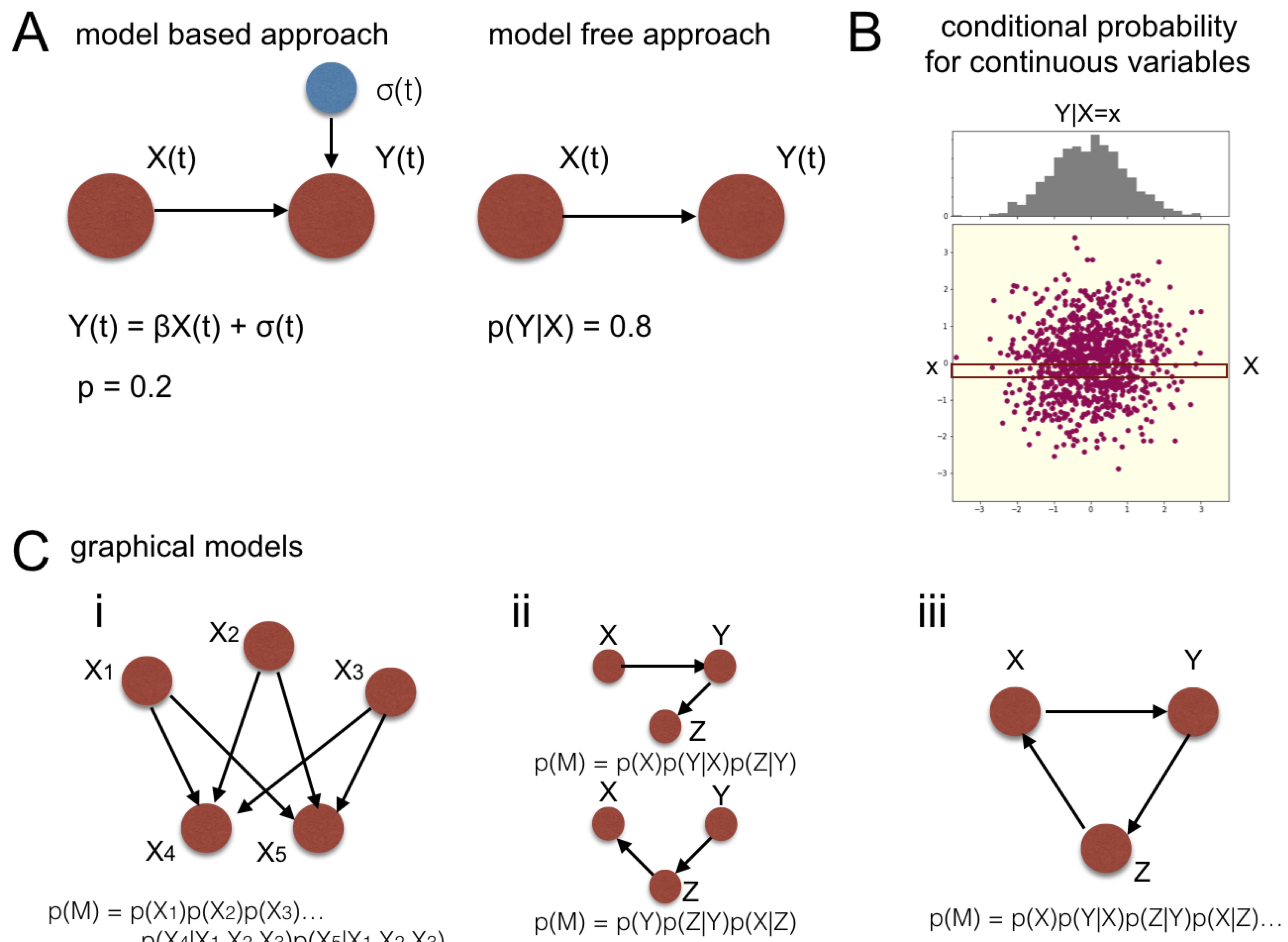}
\end{framed}
\caption{Bayesian nets. \textbf{A}: Model-based versus model-free approach. $\beta$: a~regressor coefficient fitted in the modeling procedure. $\sigma(t)$: additive noise. Both model-based and model-free approach contains a~maesure of confidence. In a~model-based approach, a~model is fitted to the~data, and~p-values associated with this fit are a~measure of confidence that the~causal link exists (i.e., is a~true positive, left panel). In a~model-free approach, this confidence is quantified directly by expressing causal relationships in terms of conditional probabilities (right panel). \textbf{B}: conditional probability for continuous variables. Since BOLD fMRI is a~continuous variables, the~joint probability distribution for variables $X$ and $Y$ is a~two-dimensional distribution. Therefore, conditional probability of $P(Y|X=x)$ becomes a~distribution. \textbf{C}: \textbf{(i)} an~exemplary Bayesian net. $X_1$, $X_2$, $X_3$: parents, $X_4$, $X_5$: children. \textbf{(ii)} competitive Bayesian nets: one can define competitive models (causal structures) in the~network and compare their~joint probability derived from the~data. \textbf{(iii)} cyclic belief propagation: if there was a~cycle in the~network, the~expression for the~joint probability would convert into an~infinite series of conditional probabilities.}
\label{fig:fig4}
\end{figure}

Then, once the whole graph is factorized into the chain of marginal and conditional probabilities, the \textit{joint probability} of the model can be computed as the product of all marginal and conditional probabilities. For instance, in Fig.~4C, (i), the~joint probability of the~model $M$ yields
\begin{equation}
P(M) = P(X_1)P(X_2)P(X_3)P(X_4|X_1, X_2, X_3)P(X_5|X_1, X_2, X_3)
\end{equation}

Finally, there are at least three possible approaches to causal inference with BNs:

\begin{enumerate}
\item model comparison: choosing the scope of possible models (by defining their structure a~priori), and comparing their joint probability. Mind that in this case, the~algorithm will simply return the~winning graphical model, without estimation of the~coefficients representing connection weights
\item assuming one model structure a~priori, and only inferring the weights. This is common practice, related to e.g. Naive Bayes~\cite{bishop2006} in which the structure is assumed, and the connectivity weights are estimated from conditional probabilities. In this case, the~algorithm will assume that the proposed graphical model is correct, and infer the~connection weights only
\item inferring the structure of the model from the data in an~iterative way, by using a~variety of approximate inference techniques that attempt to maximize posterior probability of the model by minimizing a~cost function called free energy (\cite{frey2005}, similar to DCM): expectation maximization (EM,~\cite{dempster1977,bishop2006}), variational procedures~\cite{jordan1998}, Gibbs sampling~\cite{neal1993} or the sum-product algorithm~\cite{kschischang2001} (which gives a~broader selection of procedures than in the~DCM)
\end{enumerate}

% ---------------------------------------------------------------
% Discussing of the Causation List 
% Signum of connections:
BNs can detect both excitatory and inhibitory connections $X \rightarrow Y$, depending on whether the~conditional probability $p(Y|X)$ is higher or lower than the~marginal probability $p(X)$$_1$.
% Bidirectionality:
Like LiNGAM, in general, BNs cannot pick up on bidirectional connections. The~assumption of acyclicity comes from the~cyclic belief propagation (Fig. 4C, (iii)): the~joint probability of a~cyclic graph would be expressed by an~infinite chain of conditional probabilities which usually does not converge into a~closed form. In general, this restricts the~scope of possible models to Directed Acyclic Graphs (DAGs~\cite{thulasiraman1992}). However, there are also implementations of BNs that cope with cyclic propagation of information throughout the~network, e.g. Cyclic Causal Discovery algorithm (CCD,~\cite{richardson2001}). This algorithm is not often used in practice though as it works in the~large sample limit, requires assumption on the~graph structure and retrieves a~complex output$_{4}$.
% Strength of connections, confidence intervals and resilience to confounds:
The~value of conditional probability $P(Y|X)$ can be a~measure of a~connection strength$_{2}$. We can consider conditional probabilities significantly higher than chance as an~indication for significant connections$_{3}$. In principle, BNs are not resilient to latent confounds. However, some classes of algorithms were designed especially to tackle this problem, such as Stimulus-based Causal Inference, SCI~\cite{moritzwentrup2016}, Fast Causal Inference (FCI,~\cite{spirtes1993,zhang2008}) and Greedy Fast Causal Inference (GFCI,~\cite{ogarrio2016})$_{6}$. 
% Type of inference and computational cost:
BNs can either work through model comparison or as an~exploratory technique$_7$. In the first case, it involves model specification which - like in DCM - requires a~priori knowledge about the experimental paradigm. In the~latter case, the~likelihood is intractable and can only be approximated$_8$~\cite{diggle1984}. 
% Network size:
In principle, networks of any size can be modeled with BNs, either through a~model comparison or through exploratory techniques. However, the~exploratory techniques typically minimize a~cost function during the~iterative search for the~best model. Since together with the~growing network size, the~landscape of the~cost function becomes multidimensional and complex, the~algorithm is more likely to fall into a~local minimum$_9$. 

What can also become an~issue while using BNs in practice, is the~fact that multiple BN algorithms return an~\textit{equivalence class} of a~graph: the~set of all graphs that are indistinguishable from
the~true causal structure on the~basis of their sole probabilistic independences~\cite{spirtes2010}. These structures cannot be further distinguished without further assumptions or experimental interventions. For finite data, taking even one wrong assumption upon the~directionality of causal link in the~graph can be propagated through the~network, and cause an~avalanche of incorrect orientations~\cite{spirtes2010}. One approach designed to overcome this issue is the~Constraint-Based Causal Inference (BCCD, \cite{claassen2012}). In this approach, Bayesian Inference is employed to estimate the~reliability of a~set of constraints. This estimation can further be used to decide whether this prior information should be used to determine the~causal structure in the~graph. 
% --------------------------------------------------------------

BNs cope well with noisy datasets, which makes them an~attractive option for causal research in fMRI~\cite{mumford2014}. Smith et al.~\cite{smith2011} tested multiple implementations of Bayesian nets, including FCI, CCD, as well as other algorithms: Greedy Equivalence Search (GES,~\cite{meek1995, chickering2002}), `Peter and Clark' algorithm (PC,~\cite{meek1995}) and a~conservative version of 'Peter and Clark' (CPC,~\cite{ramsey2006}). All these implementation performed similarly, which was quite well with respect to estimating the existence of connections, but not to the~directionality of the connections.

BNs are not widely used in fMRI research up to date, the~main reason being the~assumption of acyclicity. One exception is Fast Greedy Equivalence Search (FGES,~\cite{ramsey2014,ramsey2015,ramsey2017}), a~variant of GES optimized to large graphs. The~algorithm assumes that the~network is acyclic with no hidden confounders, and returns an~equivalence class for the~graph. In a~recent work by Dubois et al.~\cite{dubois2017}, FGES was applied with use of a~new, computational-experimental approach to causal inference from fMRI datasets. In the~initial step, causal inference is performed from large observational resting-state fMRI datasets with use of FGES in order to get the~aforementioned class of candidate causal structures. Further steps involve causal inference in a~single patient informed by the~results of the~initial analysis, and interventional study with use of an~electrical stimulation in order to determine which of the~equivalent structures revealed by FGES can be associated with a~particular subject.

%Independently from the aforementioned obstacles, we can also consider practicality of DAGs in terms of data requirements. Most of the DAGs fit the conditional probabilities as 2D Gaussian distributions. How many bins in a~histogram does one need to reliably fit a~2D Gaussian? Assuming as little as ten bins per one dimension (which is an~underestimation), fitting a~Gaussian conditioned on another Gaussian would require a~grid of 10 by 10, which equals to 100 bins. For a~proper estimation of any statistic, one needs roughly the amount of data per bin comparable to the amount of bins, a~hundred in this case. That altogether gives 10,000 samples, which is much longer than the best HCP resting state protocols to date (4,800 in total per subject in the newest HCP datasets~\cite{vanessen2013}). In all, independent from conceptual incompatibility with the brain research, fitting two-dimensional Gaussian functions is too data-demanding even for best available fMRI datasets.
%If we have a~large and noisy dataset, we can exchange long time series with a~small set of marginal and conditional probabilities, which make graphical models a~good method for data compression. 
%
% [c] pairwise
\section{Pairwise inference}
The~last group of methods reflects the~most recent trends in the~field of causal inference in fMRI. This family of methods is represented by Pairwise Likelihood Ratios~\cite{hyvarinen2013}, and involves a~two-stage inference procedure. In the~first step, functional connectivity is used to find connections, without assessing their directionality. Unlike network-wise methods which eliminate insignificant connections post-hoc, pairwise methods eliminate insignificant connections prior to causal inference. In the~second step, each previously found connection is analyzed separately, and the~two nodes involved are classified as an~upstream or downstream region. These methods do not involve assumptions on the~global patterns of connectivity at the~network level (recurrent versus feed-forward). However, they involve the~assumption that the~connections are \textit{non-transitive}: if $X$ projects to $Y$, and $Y$ projects to $Z$, it does not imply that $X$ projects to $Z$. The~causal inference is based on the~pairs of nodes only, and this has consequences for the interpretation of the network as a~whole. As there is uncertainty associated with estimation of every single causal link, the~probability that \emph{all} connections are correctly estimated decreases rapidly with the number of nodes in the network.
\subsection{Pairwise Likelihood Ratios}
A~two step procedure to causal inference in fMRI was first proposed by Patel (as Patel's tau, PT~\cite{patel2006}). %In this approach, the process of finding causal relations is reduced to a~binary classification problem by identifying the~(undirected) connections by means of functional connectivity. 
The first step involves identifying the~(undirected) connections by means of functional connectivity, and is achieved on the~basis of correlations between the~time series in different regions. This step results in a~binary graph of connections, and the~edges identified as empty are disregarded from further considerations, because if there is no correlation, there is no causation.

The second step determines the directionality in each one of the previously detected connections. The causal inference boils down to a~two-node Bayesian network as the whole concept is based on a~simple observation: if there is a~causal link $X \rightarrow Y$, $Y$ should get a~transient boost of activity every time $X$ increases activity. And vice versa: if there is a~causal link $Y \rightarrow X$, $X$ should react to the activation in $Y$ by increasing activity. Therefore, one can threshold the signals $X(t)$, $Y(t)$, and compute~the difference between conditional probabilities $P(Y|X)$ and $P(X|Y)$. Three scenarios are possible:
\begin{enumerate}
\item $P(Y|X)$ equals $P(X|Y)$: it is a~bidirectional connection $X \leftrightarrow Y$ (since empty connections were sorted out in the previous step)
\item the~difference between $P(Y|X)$ and $P(X|Y)$ is positive: the~connection $X \rightarrow Y$ is likely
\item the~difference between $P(Y|X)$ and $P(X|Y)$ is negative: the~connection $Y \rightarrow X$ is likely
\end{enumerate}

Building on the concept of PT, the~Pairwise Likelihood Ratios methodology (PW-LR~\cite{hyvarinen2013}) was proposed. The~authors improved on the second step of the inference by analytically deriving a~classifier to distinguish between two causal models $X \rightarrow Y$ and $Y \rightarrow X$, which corresponds to the LiNGAM model for two variables. The~authors compared the~likelihood of these two competitive models derived under LiNGAM's assumptions~\cite{hyvarinen2010}, and provided with a~cumulant based approximation to their ratio. In particular, the~authors focused on the~approximation of the likelihood ratios with third cumulant for variables $X$ and $Y$, which is an~asymmetry between first (the mean) and second (the variance) moment of the distributions of variables $X$ and $Y$ (this version of the~method is referred to by the~authors as 'PW-LR skew'): 
\begin{equation}
C_{3} = \frac{1}{N}\sum_{i = 1}^{N}(X(i) Y(i)^{2} - X(i)^{2}Y(i))
\label{eq:third_cumulant}
\end{equation}
Then, if the value of this cumulant is positive, it indicates for the connection $X \rightarrow Y$, and backwards otherwise. Additionally, the authors proposed a~modified version of the~third cumulant, including a~nonlinear transformation of the signal to improve resilience against outliers in the~signal (and referred to this modified metric as 'PW-LR r skew'). Additionally, the~authors also introduced a~version based on \textit{fourth} cumulant (referred to as 'PW-LR kurtosis').

% ---------------------------------------------------------------
% Discussing of the~Causation List 
% Signum of connections and Strength of connections:
PW-LR methods cannot distinguish between excitation and inhibition$_1$, but provide with a~quantitative measure for the strength of the connection$_2$. 
% Significance testing:
The~authors recommended to test significance of PW-LR results through permutation testing~\cite{hyvarinen2013}$_3$.
% Directionality:
Following the interpretation from Patel, it is possible to distinguish between uni- and bidirectionality (since scores close to zero might indicate the~bidirectionality)$_4$. 
% Immediacy:
The authors proposed using partial correlation instead of Pearson's correlation in the first step of the causal inference, which aims to find direct connections in the network$_5$. 
% Resilience to confounds:
As for the resilience to confounds, PW-LR methods were tested on benchmark data for which common inputs to the nodes of the network were introduced (\cite{smith2011}, simulation no 12). PW-LR gave much better performance than the best competitors (LiNGAM-ICA and PT) and reached as much as 84\% of correctly classified connections across all the benchmark data$_6$. 
% Computational cost, and type of inference:
In the~original formulation, PW-LR involves a~point estimate for the~strength of effective connectivity, and lacks estimation of confidence intervals. In such cases, in fMRI studies, estimating confidence intervals is performed in a~data-driven fashion. This is typically achieved by means of permutation testing~\cite{hyvarinen2013,smith2011} (but can also be approached with use of mixture modeling,~\cite{bielczyk2018}$_7$). PW-LR, as a~closed form solution, is computationally cheap$_8$.
As the pair-by-pair inferences do not require network fitting procedures, this can easily be applied to larger networks$_9$.
% ---------------------------------------------------------------

On the~benchmark datasets, all versions of PW-LR were performing very well, as contrasted with the~best competitors: PT and LiNGAM (and, PW-LR r skew' was giving the~best results). In all but one out of 28 simulations PW-LR was performing highly above chance, and in a~few cases they even reached 100\% accuracy. However, PW-LR has never been validated on the real fMRI datasets.

% new directions
\section{New directions in causal research in fMRI}
A number of methods have been discussed, but the~search for new ways of extracting causal information from fMRI data is still on, of which we want to highlight four representatives. First of all, one can introduce more prior knowledge into the equation. This is done in laminar analysis, where the layered structure of the cortex is assumed to contain information about the signal. Another new option is a~recently presented method based on fractional cumulants of the~BOLD distribution~\cite{bielczyk2016a}, in which the statistical properties of the BOLD fMRI signal are used for inferring causal links.

\subsection{Laminar analysis}
Advancements in fMRI acquisition have made it possible to scan at submillimetre resolution, which opens up the possibility of a~layer specific examination of the BOLD signal. As the different layers of the cortex receive and process feed forward and feedback information largely in different layers~(\cite[e.g.]{felleman1991,bastos2015}, these different processes could be visible in the laminar BOLD response. In rat studies, the BOLD response was indeed shown to have laminar specificity and have its onset in the input layer of rat motor and somatosensory cortex~\cite{Yu2014}. And also in humans, several studies suggest laminar specificity of feedback processes~\cite{Kok2016,Muckli2015}.

These results suggest that human laminar BOLD signal may contain directional and causal information. Hitherto, only single region laminar fMRI has been employed, but it may well be worthwhile to investigate how output layers of one region influence the input layer of the other. 

\subsection{Fractional cumulants}
Certain new methods take a~more statistical approach to neuroimaging data. For instance, characterizing the~shape of BOLD distributions by means of fractional moments of the~BOLD distribution combined into cumulants~\cite{bielczyk2016a} can improve on the~classification of the~two nodes within one connection into an~upstream and a~downstream node. Fractional moments of a~distribution are a~mathematical concept with limited practical interpretation, but could still contain valuable (causal) information. 

In this method a~classification procedure using fractional cumulants derived from BOLD distribution is developed. The~classifier is informed by the~DCM generative model. The initial results show that the causal classification scores similarly or better than competitive methods when applied to low-noise benchmark synthetic datasets~\cite{smith2011}, and its performance is, in general, similar to 'PW-LW r-skew'. However, the~difference shows up after imposing higher level neuronal noise on the~network: the~fractional cumulant-based classifier is the~most robust approach in presence of such natural confounds. However, validation on real fMRI datasets for this method is still pending.

\subsection{Rendering whole-brain effective connectivity with use of covariance matrices}
Recent approach to causal inference in fMRI involves inferring directionality of information transfer by using a~set of covariance matrices with both zero and nonzero time lags~\cite{gilson2016}. The~authors build a~dynamic model of the~brain network and optimize the~effective connectivity (adjacency matrix) such that the~model covariances reproduce the empirical fMRI/BOLD covariance matrices. In this way, the fitted model best matches the BOLD dynamics with respect to the second-order statistics. The~authors validate the~model in synthetic datasets, and apply to experimental fMRI datasets, using diffusion-weighted MRI imaging in order to constrain the~network connectivity. The~concept of lagged covariance matrices was also used to evaluate the~difference in cortical activation between two behavioral conditions (in application to movie watching~\cite{gilson2017}).

As this method incorporates lags, it has similar limitations as other lagged methods (such as GC or TE): it becomes lag-dependent. The~authors theoretically demonstrate that for accuracy of the~directed connectivity estimation, time lag must be matched with the time constant of the~underlying dynamical system representing the~network. How to achieve the~accuracy in order to fulfill this requirement in practice, remains an~open research question. 

Another recent contribution in this domain by Schiefer et al.~\cite{schiefer2018} focuses on inferring causal connections from resting state fMRI datasets (and other continuous time series coming from non-interventional studies), based on the assumption that the symmetric, non-lagged covariance matrix derived from the observed activity contains footprints of the direction and the sign of sparse directed connections. This underlying sparse structure is found via L1-minimization with a gradient descent, which allows for obtaining asymmetric output connectivity matrix from the initial symmetric covariance structure. In the process, the method utilizes the fact that in case of a collider present in the network ($X$ and $Y$ projecting to the same node $Z$), projecting nodes $X$ and $Y$ have a positive covariance which indicates for a particular motif in the covariance structure. The validation on ground truth synthetic datasets derived from a simple Ornstein-Uhlenbeck resulted in impressive results. On the other hand, application to the experimental fMRI datasets brought more vague results, therefore, the method requires more exploration in the fMRI datasets.

\subsection{Neural Network Models}
Another recent development relevant to the problem of causal inference is the approach of implementing neural-network models to perform a complex task that is emblematic of human cognition (most commonly visual object recognition). It is then possible to study the functional architecture and representational space of such models and attempt to draw insight from optimal model parameters as to how such tasks are implemented in the human brain. In recent years neural-network models designed to recognise objects have reached human levels of performance ~\cite{krizhevsky2012, kriegeskorte2015} and the potential of using these as models of how biological brains represent object space became a realisable goal. Early studies of feed-forward neural networks that has been replicated across multiple studies is that the closer the representational space a model uses resembles inferior temporal cortex fMRI activity the better the model performs ~\cite{yamins2013, yamins2014,khalighrazavi2014}. Of particular interest is the finding that object representations in neural-network models correlate with human brain representations in a hierarchical fashion, a result shown in across both spatial and temporal dimensions ~\cite{cichy2016}. While care must be taken not to over-interpret the generalisability of such models, these promising findings indicate that neural-network models may be able to provide insight into the fundamental constraints of certain computational processes which in turn can be applied to determining functional (and casual) relationships in human cognition.

% summary
\section{Summary}
We sum up the characteristics of all the discussed methods in the following table:

\begin{table}[ht!]
\resizebox{\textwidth}{!}
{\begin{tabular}{| l || l | l | l | l | l | l | l | l | l}
  \hline			
  Feature | Method & GC & SEM & DCM & LN & BN & TE & PW-LR\\ \hline \hline
    group of methods & net & net & net & dag & dag & net & pw \\ \hline
    sign of connections & + & + & + & + & - & + & - \\ \hline
    directionality & + & + & + & - & - & + & + \\ \hline
    connection strength & + & + & + & + & + & + & + \\ \hline
    immediacy & +/- & +/- & - & + & + & +/- & + \\ \hline     
    resilience to confounds & +/- & +/- & - & +/- & +/- & +/- & +\\ \hline
    causality through... & c & mc/c & mc & ml+c & mc/ml & c & c\\ \hline
    computational cost & l & l/h & h & h & l/h & l & l\\ \hline
    model-free? & - & - & - & - & + & + & + \\ \hline
    prespecify the graph? & - & - & + & - & +/- & - & -\\ \hline     
    regression in time & + & - & - & - & - & + & - \\ \hline
    %promising due to~\cite{smith2011}? & - & n/a & n/a & + & - & - & n/a\\ \hline
\end{tabular}}
\caption{Summary for all the methods discussed in this paper. \textit{GC}: Granger causality, \textit{SEM}: Structural Equation Modeling, \textit{DC}: Dynamic Causal Modeling, \textit{LN}: LINGaM, \textit{BN}: Bayesian nets, \textit{TE}: Transfer Entropy, \textit{PW-LR}: Pairwise Likelihood Ratios, \textit{net}: network-wise, \textit{dag}: Directed Acyclic Graphs only, \textit{pw}: pairwise, \textit{+/-}: depends on implementation, \textit{mc}: model comparison, \textit{c}: classical hypothesis testing, \textit{ml}: machine learning, \textit{l}: low, \textit{h}: high, \textit{n/a}: non-applicable. PW-LR is based on the~same concept as Patel's tau (PT), and the inference is the same, therefore we did not add a~separate column for PT.}
\label{tab:summary}
\end{table}

\section{Discussion}
% Full-network vs. DAG
In this work, we focused on discussing methods with respect to the~causal structure imposed on the~brain. According to this criterion, the~methods fall into three categories. Network-wise methods, such as GC or SEM, do not restrict the~connectivity patterns whereas Directed Acyclic Graphs (DAGs), such as BNs, assume a hierarchical structure and unidirectional connections. In the latter category, a primary node receives input from outside the~network and distributes information downstream throughout the~network. This may be a good approximation for many processes, (see for instance recent work on the visual cortex~\cite{michalareas2016}). However, the feed-forward structure assumes a~strictly hierarchical organization, which limits its capacity to model communication between different brain networks. Under what circumstances DAGs can be an accurate representation for causal structures in the~brain, remains an~open question. 

% Pairwise
Next to network-wise methods and DAGs, we also discussed a~third group of methods, referred to as `pairwise'. In this approach, the causal inference is done by splitting the inference into many pairwise inferences. Prior to this, the dimensionality is reduced based on functional connectivity, based on the~idea that (partial) correlation is a~good indicator for the~existence of causal links~\cite{smith2011} and therefore allows for simplifying the~problem, both computationally and conceptually. Since the inference in this class of methods is split into a~set of pairwise inferences, it is important to be aware of the fact that the~confidence levels are also obtained connection by connection. Therefore, for a~network represented by a~set of connections with p-values $p_i$, the joint probability of the model is roughly $\Pi_i(1-p_i)$ (in practice, confidence values for the existence of single connections are not independent, therefore this is only a~rough approximation of the~joint probability). This also means that there is a~trade-off between the~joint probability of the~graph and its density: the~joint probability of the~whole network pattern can be increased by decreasing the threshold for connectivity at more conservative p-values. Furthermore, one can look at the~pairwise inference methods as a~sort of model comparison, because in the~second step of the inference, for every connection only three options are possible to choose from. The difference with DCM procedure lies in the~fact that pairwise inference methods are based on the simple statistical properties emerging from causation in~linear systems, and do not involve minimizing the~cost function — such as negative free energy — as is done in DCM. 

%It should be noted that even successful decomposition of the~causal structure of the~brain into pairwise interactions would not fully explain 'how the brain works', the~same as analyzing the~trajectory of all bees in a~swarm would not fully explain how the~swarm operates. The brain seems to be an~orchestra without a~conductor, whose different sections dynamically synchronize, or reorganize, once the organism needs to react to the~changes in the~environment. 
% It physically hurts reading this paragraph for so many reasons (swarms do not function the way brains do, if you explicate every trajectory of all the bee's you do have an accurate description of movement of the swarm, and third the conductor is a metaphor for another issue (the absence of a central executive center), not for behaviour being emergent on the structure. So I strongly advice cutting it alltogether.

%     2. Of all methods, the only one that is commonly used is DCM. Why? Discuss.
%         tailor-made-ness? Are its assumptions closest to the data? Most interpretable? Friston's weight?
%         Do other methods get into trouble with their assumptions? 
In the~fMRI community, the~DCM family~\cite{friston2003} is currently the~most popular approach to causal inference. This is partially because DCM was tailor-made for fMRI, and includes a~generative model based on the~biological underpinnings of the~BOLD dynamics~\cite{buxton1998}. Some of the~GC studies also involve estimation of the~haemodynamic response function, and deconvolving the data before applying the~estimation procedure~\cite{david2008,ryali2011,ryali2016,hutcheson2015,wheelock2014,sathian2013,goodyear2016}. This notion of the~haemodynamics is both a~strength and a~weakness: the generative model fits the~data well, but only as long as the~current state of knowledge is accurate. New studies suggest that human haemodynamics are very dynamic and driven by state-dependent processes~\cite{miezin2000,handwerker2012}. The influence of this complex behavior on the performance of DCM is hard to estimate.

%     3. Shift towards other type of methods. Why?
%         If there is a shift towards pairwise methods, then in a discussion I want to get an idea of why that could be. For example: people couldn't resolve computationally feasible network approaches and pairwise is easier. Or, people started realising that distributional changes may be more informative than temporal forward models. I don't really know the reason. Discuss.

The~DCM procedure performs causal inference through model comparison, and as such, it is restricted to causal research in small networks containing a~few nodes - since the computational costs increase like a~factorial with the number of nodes. With the rise of research into resting state networks that contain up to 200 nodes, this may prove to be a limiting characteristic~\cite{smith2009}. This issue can be addressed with new methods for pairwise inference such as PT and PW-LR, which do not impose any upper bound on the size of the~network. 

%     5. Assumptions. Not sure what you'd like here. I'm inclined to discuss that in relation to method and a bit earlier.
It is important to remember that there are always two aspects to a~method for causal inference. First, the method should have assumptions grounded in a~biologically plausible framework, well-suited for the given dataset. For instance, a~method for causal inference in fMRI should respect: (1) the confounding, region- and subject-specific BOLD dynamics~\cite{handwerker2004}; and (2) co-occurance of cause and effect (since the~time resolution of the data is low compared to the~underlying neuronal dynamics, the~causes and their effects most likely happen within the~same frame in the~fMRI data). The~new methods for pairwise inference address this issue by (1) breaking the~time order, and performing causal inference on the~basis of statistical properties of the~distribution of the BOLD samples, and not from the~timing of events; (2) using correlation in order to detect connections. A~good counterexample here is GC. GC has been proven useful in multiple disciplines, and its estimation procedure is impeccable: nonparametric, computationally straightforward, and it gives a~unique, unbiased solution. However, there is an ongoing~discussion on whether or not GC is suited for causal interpretations of fMRI data. On the one hand, theoretical work by Seth et al.~\cite{seth2013} and Roebroeck et al.~\cite{roebroeck2005} suggest that despite the~slow haemodynamics, GC can still be informative about the~directionality of causal links in the~brain. On the~other hand, the~work by Webb~\cite{webb2013} demonstrates that the spatial distribution of GC corresponds to the Circle of Willis, the major blood vessels in the~brain. 

%     6. Estimation method
Secondly, an~estimation procedure needs to be computationally stable. Even if the~generative model faithfully describes the data, it still depends on the estimation algorithm whether the method will return \textit{correct} results. However, the face-validity of the~algorithms can only be tested in particular paradigms, in which the~ground truth is known. If in the~given paradigm, the~ground truth is unknown - which is most often the~case in fMRI experiments - only reliability can be tested. One way of assessing reliability of the~method is testing for the~test-retest convergence. So far, DCM is %a~good example here: the~generative model was based on detailed knowledge of the physiology of the brain, and as such, it is currently the best representation of BOLD fMRI as a~function of network dynamics. It is also 
the only method that has been extensively tested in terms of test-retest reliability in separate studies~\cite{frassle2015,frassle2016a,schuyler2010,rowe2010,tak2018}, and performed good overall. In general, it is desirable to have more studies testing the~reliability of the methods on reliability in experimental fMRI datasets - as such validation of multiple methods such as GC or SEM, is still missing.%The~estimation procedure can be stochastic (in the MCMC version) or deterministic (VB version). 

One last remark about the nature of the different methods: some methods are developed for event-related fMRI, such as DCM. Yet, new implementations of spectral DCM for the resting state as well~\cite{friston2014}. As for other methods, application to resting-state studies is relatively straightforward, while task fMRI can pose certain constraints on the~methods. For instance, lag-based methods such as GC work best when the~task is executed in a~form of epochs~\cite{deshpande2008a} rather than a~few second stimulus-response blocks, because it is extremely difficult to fit an AR model to datasets of 1-2 frames in length. For this reason, structural methods (which do not regard the~time sequence) such as BNs or PWLR, will be much more efficient in estimating causality in such cases.

Coming back to the main question posed in this review, can we hope to uncover causal relations in the~brain using fMRI? Although there are new concepts in the~field, which propose to consider causal interactions in the~brain in probabilistic terms~\cite{mannino2015,griffiths2015}, the~'traditional', deterministic models of causality are prevalent in neuroimaging. Within these deterministic models, in the~light of the existing literature, the new research directions based on breaking the time order as the~axiom of causal inference (such as PWLR, PT, and LiNGAM), prove more successful than the more 'traditional' approaches which take regression in time into account (such as GC or TE,~\cite{smith2011,hyvarinen2013}). Also, Patel's two-step design to achieve a~causal map of connections is very promising, especially once the~Pearson correlation is replaced with partial correlation as is done in PW-LR. One note to add is that 'success' of any method for causal inference in fMRI depends on the~forward model used for generating the~synthetic dataset. In the~seminal paper by Smith et al. we are referring to,~\cite{smith2011}, multiple methods were evaluated and critically discussed on the~basis of simulations of the DCM generative model. However, there are alternatives, e.g., generative model by Seth et al.~\cite{seth2013}, which might potentially yield other hierarchy of methods in terms of success rate.

In this paper, we discuss the topic of inferring causal processes from fMRI datasets on the individual subject. One approach that could further contribute to the~development of methods for causal inference in fMRI though, is a~\textit{group inference} approach. In such an~approach, a~prior that different subjects represent similar causal structures, is added to the~inference procedure. As lumping the~datasets coming from different subjects increases the~amount of data to derive the~causal structure from, this assumption, in general, facilitates the~inference. Multiple algorithms for group inference for effective connectivity in fMRI have already been proposed, including Independent Multiple sample Greedy Equivalence Search (IMaGES,~\cite{ramsey2010}), LOFS algorithm previously mentioned in chapter~\ref{sec:lingam}~\cite{ramsey2011} and Group Iterative Multiple Model Estimation (GIMME,~\cite{gates2012}). 

Furthermore, with the~current rapid growth of translational research and increase in use of invasive and acute stimulation techniques such as optogenetics~\cite{deisseroth2011,ryali2016} or TMS~\cite{kim2009}, a~rigid validation of methodology for causal inference becomes feasible through interventional studies. Recently, multiple methods for inferring causality from fMRI data were validated using a joint fMRI and MEG experiment~\cite{mill2017}, with promising results for GC and BNs. This gives hope for establishing causal relations in neural networks, using fMRI.
\section*{Disclosure/Conflict-of-Interest Statement}
The authors declare that the research was conducted in the absence of any commercial or financial relationships that could be construed as a potential conflict of interest. JCG has acted as a~consultant to Boehringer Ingelheim in the last 4 years, but is not an~employee or shareholder of this company.

\section*{Author Contributions}
NZB drafted the manuscript. NZB, SU, and TvM restructured the manuscript. NZB, SU, TvM and PA revised the work as a team. JKB and JCG critically revised the final manuscript.

\section*{Acknowledgements}
We would like to thank to \textbf{Lionel Barnett, Christian Beckmann, Daniel Borek, Patrick Ebel, Daniel Gomez, Moritz Grosse-Wentrup, Max Hinne, Maciej Jedynak, Christopher Keown, S$\acute{a}$ndor Kolumb$\acute{a}$n, Vinod Kumar, Randy McIntosh, Nils M\"{u}ller, Hanneke den Ouden, Payam Piray, Thomas Rhys-Marshall, Gido Schoenmacker, Ghaith Tarawneh, Fabian Walocha} and \textbf{Johannes Wilbertz} for sharing knowledge about causal inference in fMRI, and for providing a~valuable content. We would like to further thank \textbf{Martha Nari-Havenith} and \textbf{Peter Vavra} for his contribution to the~conceptual work. In addition, we would like to cordially thank \textbf{Thomas Wolfers} for encouragement and help at an~early stage.

\section*{Funding} 
NB, MNH, JG and JB are supported by the European Community's Seventh Framework Programme (FP7/2007-2013) under grant agreement no 305697 (OPTIMISTIC), the European Community's Seventh Framework Programme (FP7/2007-2013) under grant agreement no 278948 (TACTICS) and European Union’s Seventh Framework Programme for research, technological development and demonstration under grant agreement no 603016 (MATRICS). SU was supported by grant 657605 of the Marie Sklodowska-Curie Horizon 2020 framework of the European Union.

\bibliography{myrefs_all}
\end{document}